\def\neutralinoa{$\widetilde{\chi}_{1}^{0}$}
\def\neutralinob{$\widetilde{\chi}_{2}^{0}$}

\def\charginoa{$\widetilde{\chi}_{1}^{\pm}$}

\def\sq{$\widetilde{q}$}

\def\gluino{$\widetilde{g}$}

\def\ttbar{$t\overline{t}$}

\def\ET{$E_{T}$}

\def\HT{$H_{T}$}
\def\MT{$M_{T}$}
\def\MET{\mbox{${\hbox{$E$\kern-0.6em\lower-.1ex\hbox{/}}}_T$}}
\def\calmet{\mbox{${\hbox{$E$\kern-0.6em\lower-.1ex\hbox{/}}}_T^{\rm cal}$}}
\def\vecmet{\mbox{$\vec{{\hbox{$E$\kern-0.6em\lower-.1ex\hbox{/}}}}_T$}}

\def\Nqcd{$N_{\rm multijet}$}
\def\Ntt{$N_{t\overline{t}}$}
\def\etal{{\sl et al.}}

\documentstyle[prd,epsf,epsfig,eqsecnum,aps,floats]{revtex} 

\author{                                                                      
V.M.~Abazov,$^{23}$                                                           
B.~Abbott,$^{57}$                                                             
A.~Abdesselam,$^{11}$                                                         
M.~Abolins,$^{50}$                                                            
V.~Abramov,$^{26}$                                                            
B.S.~Acharya,$^{17}$                                                          
D.L.~Adams,$^{55}$                                                            
M.~Adams,$^{37}$                                                              
S.N.~Ahmed,$^{21}$                                                            
G.D.~Alexeev,$^{23}$
% ...
\\                                                                            
\vskip 0.30cm                                                                 
\centerline{(D\O\ Collaboration)}                                             
\vskip 0.30cm                                                                 
}                                                                             
\address{                                                                     
\centerline{$^{1}$Universidad de Buenos Aires, Buenos Aires, Argentina}       
\centerline{$^{2}$LAFEX, Centro Brasileiro de Pesquisas F{\'\i}sicas,         
                  Rio de Janeiro, Brazil}                                     
\centerline{$^{3}$Universidade do Estado do Rio de Janeiro,                   
                  Rio de Janeiro, Brazil}                                     
\centerline{$^{4}$Institute of High Energy Physics, Beijing,                  
                  People's Republic of China}                                 
\centerline{$^{5}$Universidad de los Andes, Bogot\'{a}, Colombia}             
\centerline{$^{6}$Charles University, Center for Particle Physics,            
                  Prague, Czech Republic}                                     
\centerline{$^{7}$Institute of Physics, Academy of Sciences, Center           
                  for Particle Physics, Prague, Czech Republic}               
\centerline{$^{8}$Universidad San Francisco de Quito, Quito, Ecuador}         
\centerline{$^{9}$Institut des Sciences Nucl\'eaires, IN2P3-CNRS,             
                  Universite de Grenoble 1, Grenoble, France}                 
\centerline{$^{10}$CPPM, IN2P3-CNRS, Universit\'e de la M\'editerran\'ee,     
                  Marseille, France}                                          
\centerline{$^{11}$Laboratoire de l'Acc\'el\'erateur Lin\'eaire,              
                  IN2P3-CNRS, Orsay, France}                                  
\centerline{$^{12}$LPNHE, Universit\'es Paris VI and VII, IN2P3-CNRS,         
                  Paris, France}                                              
\centerline{$^{13}$DAPNIA/Service de Physique des Particules, CEA, Saclay,    
                  France}                                                     
\centerline{$^{14}$Universit{\"a}t Mainz, Institut f{\"u}r Physik,            
                  Mainz, Germany}                                             
\centerline{$^{15}$Panjab University, Chandigarh, India}                      
\centerline{$^{16}$Delhi University, Delhi, India}                            
\centerline{$^{17}$Tata Institute of Fundamental Research, Mumbai, India}     
\centerline{$^{18}$Seoul National University, Seoul, Korea}                   
\centerline{$^{19}$CINVESTAV, Mexico City, Mexico}                            
\centerline{$^{20}$FOM-Institute NIKHEF and University of                     
                  Amsterdam/NIKHEF, Amsterdam, The Netherlands}               
\centerline{$^{21}$University of Nijmegen/NIKHEF, Nijmegen, The               
                  Netherlands}                                                
\centerline{$^{22}$Institute of Nuclear Physics, Krak\'ow, Poland}            
\centerline{$^{23}$Joint Institute for Nuclear Research, Dubna, Russia}       
\centerline{$^{24}$Institute for Theoretical and Experimental Physics,        
                   Moscow, Russia}                                            
\centerline{$^{25}$Moscow State University, Moscow, Russia}                   
\centerline{$^{26}$Institute for High Energy Physics, Protvino, Russia}       
\centerline{$^{27}$Lancaster University, Lancaster, United Kingdom}           
\centerline{$^{28}$Imperial College, London, United Kingdom}                  
\centerline{$^{29}$University of Arizona, Tucson, Arizona 85721}              
\centerline{$^{30}$Lawrence Berkeley National Laboratory and University of    
                  California, Berkeley, California 94720}                     
\centerline{$^{31}$University of California, Davis, California 95616}         
\centerline{$^{32}$California State University, Fresno, California 93740}     
\centerline{$^{33}$University of California, Irvine, California 92697}        
\centerline{$^{34}$University of California, Riverside, California 92521}     
\centerline{$^{35}$Florida State University, Tallahassee, Florida 32306}      
\centerline{$^{36}$Fermi National Accelerator Laboratory, Batavia,            
                   Illinois 60510}                                            
\centerline{$^{37}$University of Illinois at Chicago, Chicago,                
                   Illinois 60607}                                            
\centerline{$^{38}$Northern Illinois University, DeKalb, Illinois 60115}      
\centerline{$^{39}$Northwestern University, Evanston, Illinois 60208}         
\centerline{$^{40}$Indiana University, Bloomington, Indiana 47405}            
\centerline{$^{41}$University of Notre Dame, Notre Dame, Indiana 46556}       
\centerline{$^{42}$Iowa State University, Ames, Iowa 50011}                   
\centerline{$^{43}$University of Kansas, Lawrence, Kansas 66045}              
\centerline{$^{44}$Kansas State University, Manhattan, Kansas 66506}          
\centerline{$^{45}$Louisiana Tech University, Ruston, Louisiana 71272}        
\centerline{$^{46}$University of Maryland, College Park, Maryland 20742}      
\centerline{$^{47}$Boston University, Boston, Massachusetts 02215}            
\centerline{$^{48}$Northeastern University, Boston, Massachusetts 02115}      
\centerline{$^{49}$University of Michigan, Ann Arbor, Michigan 48109}         
\centerline{$^{50}$Michigan State University, East Lansing, Michigan 48824}   
\centerline{$^{51}$University of Nebraska, Lincoln, Nebraska 68588}           
\centerline{$^{52}$Columbia University, New York, New York 10027}             
\centerline{$^{53}$University of Rochester, Rochester, New York 14627}        
\centerline{$^{54}$State University of New York, Stony Brook,                 
                   New York 11794}                                            
\centerline{$^{55}$Brookhaven National Laboratory, Upton, New York 11973}     
\centerline{$^{56}$Langston University, Langston, Oklahoma 73050}             
\centerline{$^{57}$University of Oklahoma, Norman, Oklahoma 73019}            
\centerline{$^{58}$Brown University, Providence, Rhode Island 02912}          
\centerline{$^{59}$University of Texas, Arlington, Texas 76019}               
\centerline{$^{60}$Texas A\&M University, College Station, Texas 77843}       
\centerline{$^{61}$Rice University, Houston, Texas 77005}                     
\centerline{$^{62}$University of Virginia, Charlottesville, Virginia 22901}   
\centerline{$^{63}$University of Washington, Seattle, Washington 98195}       
}                                                                             

\begin{document}

\title {Search for mSUGRA in single-electron events with jets and large
missing transverse energy in $p\overline{p}$ collisions at $\sqrt{s} =
1.8$ TeV}

\maketitle

\hspace{1.5in}
\centerline{\large Abstract}
\begin{abstract}
We describe a search for evidence of minimal supergravity (mSUGRA) in
$92.7 \; \rm{pb}^{-1}$ of data collected with the D\O\ detector at the
Fermilab Tevatron $p\overline{p}$ collider at $\sqrt{s} = 1.8 \; {\rm
TeV}$. Events with a single electron, four or more jets, and large
missing transverse energy were used in this search. The major
backgrounds are from $W$+jets, misidentified multijet, \ttbar, and $WW$
production. We observe no excess above the expected number of background
events in our data. A new limit in terms of mSUGRA model parameters is
obtained.
\end{abstract}

\pacs{PACS numbers:}


\section{INTRODUCTION} \label{sec:introduction}

The standard model (SM) has been a great achievement in particle
physics. A large number of experimental results have confirmed many
features of the theory to a high degree of precision. However, the SM is
theoretically unsatisfactory, and it poses many questions and
problems~\cite{quigg_p188,r_cahn}. The most notable ones are the
fine-tuning problem of the SM Higgs self-interaction through fermion
loops~\cite{j_hewett} and the unknown origin of electroweak symmetry
breaking (EWSB).~ Supersymmetry (SUSY)~\cite{susy_review} incorporates
an additional symmetry between fermions and bosons, and offers a
solution to the fine-tuning problem and a possible mechanism for EWSB.

SUSY postulates that for each SM degree of freedom, there is a
corresponding SUSY degree of freedom. This results in a large number of
required supersymmetric particles (sparticles), and at least two Higgs
doublets in the theory. A new quantum number, called
$R$-parity~\cite{farrar_fayet}, is used to distinguish between SM
particles and sparticles. All SM particles have $R$-parity $+1$ and
sparticles have $R$-parity $-1$. The simplest extension to the SM, the
minimal supersymmetric standard model (MSSM), respects the same
$\mbox{SU(3)} \otimes \mbox{SU(2)} \otimes \mbox{U(1)}$ gauge symmetries
as does the SM. SUSY must be a broken symmetry. Otherwise we would have
discovered supersymmetric particles of the same masses as their SM
partners. A variety of models have been proposed for SUSY breaking. One
of these, the minimal supergravity (mSUGRA) model, postulates that
gravity is the communicating force from the SUSY breaking origin at a
high mass scale to the electroweak scale, which is accessible to current
high energy colliders. The mSUGRA model is described in detail in
Ref.~\cite{run2_sugra}. It can be characterized by four parameters and a
sign: a common scalar mass ($m_{0}$), a common gaugino mass ($m_{1/2}$),
a common trilinear coupling value ($A_{0}$), the ratio of the vacuum
expectation values of the two Higgs doublets ($\tan\beta$), and the sign
of $\mu$, where $\mu$ is the Higgsino mass parameter.

In this analysis, $R$-parity is assumed to be conserved. This implies
that sparticles must be pair-produced in $p\overline{p}$ collisions. The
sparticles can decay directly, or via lighter sparticles, into final
states that contain SM particles and the lightest supersymmetric
particles (LSPs), which must be stable. Because the LSP interacts
extremely weakly, it escapes detection and leaves a large imbalance in
transverse energy (\MET) in the event. We assume that the lightest
neutralino (\neutralinoa) is the LSP, and that $A_{0}=0$ and $\mu<0$. We
fix $\tan\beta = 3$ and perform the search in the $m_{1/2}$--$m_{0}$
plane.

\begin{figure} \centering
	\epsfig{file=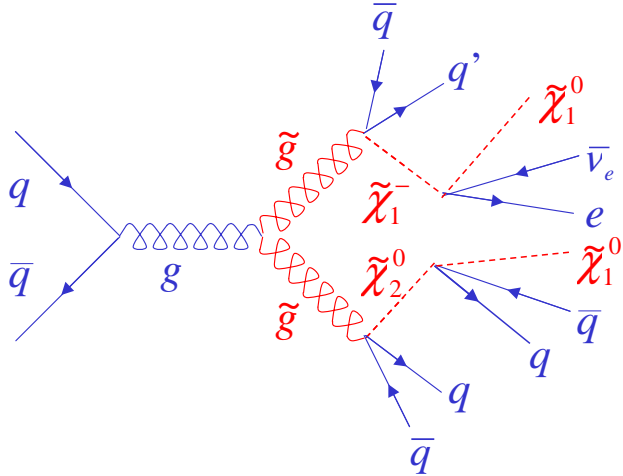, width=0.55\hsize, angle=270}
	\caption{Feynman diagram for gluino pair production and decay to
	an electron, multijets, and produce \MET. The three-body decays
	are in fact cascade decays in which off-shell particles or
	sparticles are produced.}
	\label{fig:feynman_gluino_gluino}
\end{figure}

Most recently, searches for mSUGRA signatures have been performed at LEP
and the Tevatron. At D\O, dilepton+\MET~\cite{dilepton} and
jets+\MET~\cite{jet_met} final states have been examined for possible
mSUGRA effects. This report describes a search in the final state
containing a single isolated electron, four or more jets, and large
\MET. One of the possible mSUGRA particle-production processes which
results in such a final state is shown in
Fig.~\ref{fig:feynman_gluino_gluino}. The search is particularly
sensitive to the moderate $m_{0}$ region where charginos and neutralinos
decay mostly into SM $W$ and/or $Z$ bosons which have large branching
fractions to jets. It also complements our two previous searches since
the signatures are orthogonal to one another.

\section{THE D\O\ DETECTOR} \label{sec:d0detector}

D\O\ is a multipurpose detector designed to study $p\overline{p}$
collisions at the Fermilab Tevatron Collider. The work presented here is
based on approximately $92.7 \; \rm{pb}^{-1}$ of data recorded during
the 1994--1996 collider runs. A full description of the detector can be
found in Ref.~\cite{d0nim}. Here, we describe briefly the properties of
the detector that are relevant for this analysis.

\begin{figure}
\epsfig{file=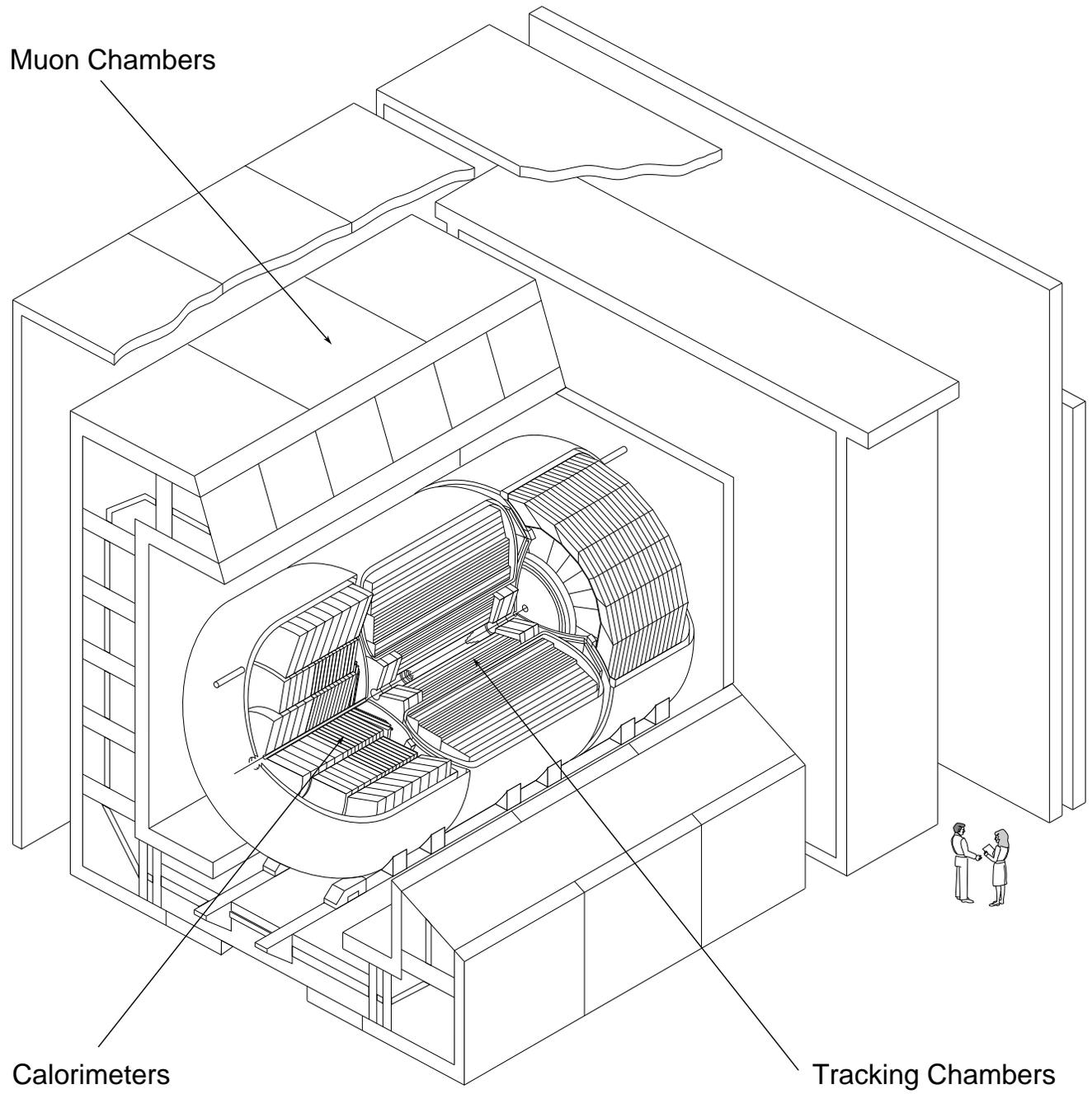,width=\hsize}
\caption{Cut away isometric view of the D\O\ detector.}
\label{fig:d0_isometric}
\end{figure}

The detector was designed to have good electron and muon identification
capabilities and to measure jets and \MET\ with good resolution.  The
detector consists of three major systems: a non-magnetic central
tracking system, a uranium/liquid-argon calorimeter, and a muon
spectrometer. A cut-away view of the detector is shown in
Fig.~\ref{fig:d0_isometric}.

The central detector (CD) consists of four tracking subsystems: a vertex
drift chamber, a transition radiation detector, a central drift chamber,
and two forward drift chambers. It measures the trajectories of charged
particles and can discriminate between singly-charged particles and
$e^+e^-$ pairs from photon conversions through the ionization measured
along their tracks. It covers the pseudorapidity~\cite{eta_definition}
region $|\eta_{\rm d}| < 3.2$.

The calorimeter is divided into three parts: the central calorimeter
(CC) and the two end calorimeters (EC), each housed in its own steel
cryostat, which together cover the pseudorapidity range $|\eta_{\rm d}|
< 4.2$. Each calorimeter consists of an inner electromagnetic (EM)
section, a fine hadronic (FH) section, and a coarse hadronic (CH)
section. Between the CC and the EC is the inter-cryostat detector (ICD),
which consists of scintillator tiles. The EM portion of the calorimeters
is 21 radiation lengths deep and is divided into four longitudinal
segments (layers). The hadronic portions are 7--9 nuclear interaction
lengths deep and are divided into four (CC) or five (EC) layers. The
calorimeters are segmented transversely into pseudoprojective towers of
$\Delta\eta\times\Delta\phi = 0.1 \times 0.1$. The third layer of the EM
calorimeter, where most of the EM shower energy is expected, is
segmented twice as finely in both $\eta$ and $\phi$, with cells of size
$\Delta\eta\times\Delta\phi$ = $0.05 \times 0.05$. The energy resolution
for electrons is $\sigma(E)/E = 15\%/\sqrt{E(\rm{GeV})} \oplus
0.4\%$. For charged pions, the resolution is $50\%/\sqrt{E(\rm{GeV})}$
and for jets $80\%/\sqrt{E(\rm{GeV})}$. The resolution in \MET\ is $1.08
\; {\rm GeV} + 0.019 \cdot \sum E_{T}({\rm GeV})$, where $\sum E_{T}$ is
the scalar sum of the transverse energies in all calorimeter cells.

The wide angle muon system (WAMUS), which covers $|\eta_{\rm d}|<2.5$,
is also used in this analysis. The system consists of four planes of
proportional drift tubes in front of magnetized iron toroids with a
magnetic field of 1.9~T and two groups of three planes of proportional
drift tubes behind the toroids. The magnetic field lines and the wires
in the drift tubes are transverse to the beam direction. The muon
momentum $p$ is measured from the muon's angular bend in the magnetic
field of the iron toroids, with a resolution of $\sigma(1/p) = 0.18(p-2
\; {\rm GeV})/p^{2} \oplus 0.003 \; {\rm GeV}^{-1}$, for $p>4.0 \;
{\rm GeV}$.

A separate synchrotron, the Main Ring, lies above the Tevatron and goes
through the CH calorimeter. During data-taking, it is used to accelerate
protons for antiproton production. Particles lost from the Main Ring can
deposit significant energy in the calorimeters, increasing the
instrumental background. We reject much of this background at the
trigger level by not accepting events during beam injection into the
Main Ring, when losses are largest.

\section{EVENT SELECTION} \label{sec:selection}

Event selection at D\O\ is performed at two levels: online selection at
the trigger level and offline selection at the analysis level. The
algorithms to reconstruct the physical objects (electron, muon, jet,
\MET) as well as their identification at the online and offline levels are
described in Ref.~\cite{run1a_top_prd}. We summarize below the
selections pertaining to this analysis.

\subsection{Triggers} \label{subsec:triggers}
The D\O\ trigger system reduces the event rate from the beam crossing
rate of 286~kHz to approximately 3--4~Hz, at which the events are
recorded on tape. For most triggers (and those we use in this analysis)
we require a coincidence in hits between the two sets of scintillation
counters located in front of each EC (level~0). The next stage of the
trigger (level~1) forms fast analog sums of the transverse energies in
calorimeter trigger towers. These towers have a size of
$\Delta\eta\times\Delta\phi = 0.2 \times 0.2$, and are segmented
longitudinally into EM and FH sections. The level~1 trigger operates on
these sums along with patterns of hits in the muon spectrometer. A
trigger decision can be made between beam crossings (unless a level 1.5
decision is required, as described below). After level~1 accepts an
event, the complete event is digitized and sent to the level~2 trigger,
which consists of a farm of 48~general-purpose processors. Software
filters running in these processors make the final trigger decision.

The triggers are defined in terms of combinations of specific objects
required in the level~1 and level~2 triggers. These elements are
summarized below. For more information, see
Refs.~\cite{d0nim,run1a_top_prd}.

To trigger on electrons, level~1 requires that the transverse energy in
the EM section of a trigger tower be above a programmed threshold. The
level~2 electron algorithm examines the regions around the level~1
towers that are above threshold, and uses the full segmentation of the
EM calorimeter to identify showers with shapes consistent with those of
electrons. The level~2 algorithm can also apply an isolation requirement
or demand that there be an associated track in the central detector.

For the later portion of the run, a ``level~1.5'' processor was also
available for electron triggering. In this processor, each EM trigger
tower above the level~1 threshold is combined with the neighboring tower
of the highest energy. The hadronic portions of these two towers are
also combined, and the ratio of EM transverse energy to total transverse
energy in the two towers is required to be $> 0.85$. The use of a
level~1.5 electron trigger is indicated in the tables below as an ``EX''
tower.

The level~1 muon trigger uses the pattern of drift tube hits to provide
the number of muon candidates in different regions of the muon
spectrometer. A level~1.5 processor can also be used to put a $p_{T}$
requirement on the candidates (at the expense of slightly increased dead
time). At level~2, the fully digitized event is available, and the first
stage of the full event reconstruction is performed. The level~2 muon
algorithm can also require the presence of energy deposition in the
calorimeter consistent with that from a muon.

For a jet trigger, level~1 requires that the sum of the transverse
energies in the EM and hadronic sections of a trigger tower be above a
programmed threshold. Level~2 then sums calorimeter cells around the
identified towers (or around the \ET-weighted centroids of the large
tiles) in cones of a specified radius $\Delta R =
\sqrt{\Delta\eta^2 + \Delta\phi^2}$, and imposes a threshold on the
total transverse energy.

The \MET\ in the calorimeter is computed both at level~1 and
level~2. For level~1, the vertex $z$ position is assumed to be at the
center of the detector, while for level~2, the vertex $z$ position is
determined from the relative timing of hits in the level~0 scintillation
counters.

The trigger requirements used for this analysis are summarized in
Table~\ref{tbl:triggers}. Runs taken during 1994--1995 (Run 1b) and
during the winter of 1995--1996 (Run 1c) were used, and only the
triggers ``ELE\_JET\_HIGH'' and ``ELE\_JET\_HIGHA'' in the table were
used to conduct this search for mSUGRA. The ``EM1\_EISTRKCC\_MS''
trigger was used for background estimation. As mentioned above, these
triggers do not accept events during beam injection into the main
ring. In addition, we do not use events which were collected when a Main
Ring bunch passed through the detector or when losses were registered in
monitors around the Main Ring. Several bad runs resulting from hardware
failure were also rejected. The ``exposure'' column in
Table~\ref{tbl:triggers} takes these factors into account.

\begin{table*}
\caption{Triggers used during Run 1b and Run 1c. ``Exposure'' gives the
effective integrated luminosity for each trigger, taking into
account the Main Ring vetoes and bad runs.}
\squeezetable
\begin{minipage}{\linewidth}
\begin{tabular}{lcccc}
Trigger Name & Exposure & Level 1 & Level 2 & Run \\
             & ($\rm{pb}^{-1}$) & & & period \\
\tableline
EM1\_EISTRKCC\_MS & 82.9 & 1 EM tower, $E_{T} > 10 \; \rm{GeV}$
                         & 1 isolated $e$, $E_{T} > 20 \; \rm{GeV}$
                         & Run 1b \\
                  &      & 1 EX tower, $E_{T} > 15 \; \rm{GeV}$
                         & $\mbox{\calmet} > 15\;\rm{GeV}$~\footnote{\calmet\ is the missing \ET\ in the calorimeter, obtained from the sum of transverse energy of all calorimeter cells. \MET\ is the missing \ET\ corrected for muon momentum, obtained by subtracting the transverse momenta of identified muons from \calmet.}
                         & \\
\tableline
                  &      & 1 EM tower, $E_{T} > 12 \; \rm{GeV}$, $|\eta| < 2.6$ 
                         & 1 $e$, $E_{T} > 15 \; \rm{GeV}$, $|\eta| < 2.5$
                         & \\
ELE\_JET\_HIGH	  & 82.9 & 2 jet towers, $E_{T} > 5 \; \rm{GeV}$, $|\eta| < 2.0$   
                         & 2 jets ($\Delta R=0.3$), $E_{T} > 10 \; \rm{GeV}$, $|\eta| < 2.5$
                         & Run 1b \\
		  &	 & & $\mbox{\calmet} > 14 \;\rm{GeV}$ \\
\tableline
ELE\_JET\_HIGH	  & 0.89 & ditto & ditto & Run 1c \\
\tableline
                  &      & 1 EM tower, $E_{T} > 12 \; \rm{GeV}$, $|\eta| < 2.6$
                         & 1 $e$, $E_{T} > 17 \; \rm{GeV}$, $|\eta| < 2.5$
                         & \\
ELE\_JET\_HIGHA   & 8.92 & 2 jet towers, $E_{T} > 5 \; \rm{GeV}$, $|\eta| < 2.0$   
                         & 2 jets ($\Delta R=0.3$), $E_{T} > 10 \; \rm{GeV}$, $|\eta| < 2.5$
                         & Run 1c \\
		  &	 & & $\mbox{\calmet} > 14 \;\rm{GeV}$ \\
\end{tabular}
\end{minipage}
\label{tbl:triggers}
\end{table*}

\subsection{Object Identification} \label{object_id}

\subsubsection{Electrons}
Electron identification is based on a likelihood technique. Candidates
are first identified by finding isolated clusters of energy in the EM
calorimeter with a matching track in the central detector. We then cut
on a likelihood constructed from the following five variables:

\begin{itemize}
\item a $\chi^{2}$ from a covariance matrix that checks the
consistency of the shape of a calorimeter cluster with that expected of
an electron shower;
\item an electromagnetic energy fraction, defined as the ratio of the
portion of the energy of the cluster found in the EM calorimeter to its
total energy;
\item a measure of consistency between the trajectory in the tracking
chambers and the centroid of energy cluster (track match significance);
\item the ionization deposited along the track $dE/dx$;
\item a measure of the radiation pattern observed in the transition
radiation detector (TRD). (This variable is used only for CC EM clusters
because the TRD does not cover the forward region~\cite{d0nim}.)
\end{itemize}

\noindent To a good approximation, these five variables are independent
of each other.

High energy electrons in mSUGRA events tend to be isolated. Thus, we use
the additional restriction:

\begin{equation}
{E_{\rm{tot}}(0.4) - E_{\rm{EM}}(0.2) \over E_{\rm{EM}}(0.2)} < 0.1,
\end{equation}

\noindent where $E_{\rm{tot}}(0.4)$ is the energy within $\Delta R <
0.4$ of the cluster centroid ($\Delta R = \sqrt{\Delta\eta^2 +
\Delta\phi^2}$) and $E_{\rm{EM}}(0.2)$ is the energy in the EM
calorimeter within $\Delta R < 0.2$. We denote this restriction the
``isolation requirement.''

The electron identification efficiency, $\varepsilon^{e}_{\rm id}$, is
measured using the $Z \rightarrow ee$ data. Since only CC
($|\eta^{e}_{\rm d}|<1.1$) and EC ($1.5<|\eta^{e}_{\rm d}|<2.5$) regions
are covered by EM modules, electron candidates are selected and their
identification efficiencies are measured in these two regions. An
electron is considered a ``probe'' electron if the other electron in the
event passes a strict likelihood requirement. This gives a clean and
unbiased sample of electrons. We construct the invariant mass spectrum
of the two electron candidates and calculate the number of background
events, which mostly come from Drell-Yan production and misidentified
jets, inside a $Z$ boson mass window. After background subtraction, the
ratio of the number of events inside the $Z$ boson mass window before
and after applying the likelihood and isolation requirements to each
probe electron, gives $\varepsilon^{e}_{\rm id}$.

The $\varepsilon^{e}_{\rm id}$ is a function of jet multiplicity in the
event. The presence of jets reduces $\varepsilon^{e}_{\rm id}$ primarily
due to the isolation requirement and reduced tracking
efficiency. However, with a larger numbers of jets ($\geq 3$) in the
event, the efficiency of locating the correct hard-scattering vertex
increases. The two effects compensate each other for events with high
jet multiplicity~\cite{zhou_thesis}. The electron identification
efficiencies used in this analysis are obtained from $Z \rightarrow ee$
data with at least two jets and are given in Table~\ref{tbl:eid_eff}.

Sometimes a jet with very similar characteristics to an electron can
pass the electron identification selection, and result in a fake
electron. The effect of fake electrons is discussed in
section~\ref{subsec:qcd_bkgd}.

\subsubsection{Jets}
Jets are reconstructed in the calorimeter using a fixed-size cone
algorithm with $\Delta R = 0.5$. A jet that originates from a quark or a
gluon deposits a large fraction of its energy in the FH part of the
calorimeter, and so we identify jets through the fractional energy in
the EM and CH parts of the calorimeter. We require the fraction of the
total jet energy deposited in the EM section of the calorimeter ({\it
emf}\hspace{1.0mm}) to be between 0.05 and 0.95 for high energy jets
($E^{j}_{T}>35 \; {\rm GeV}$), and the fraction of the total jet energy
deposited in the CH section of the calorimeter ({\it chf}\hspace{1.0mm})
to be less than 0.4. Because electronic and uranium noise is generally
of low energy, the lower bound of the {\it emf}\hspace{0.5mm}
requirement is raised gradually for lower energy jets in the CC. (It is
0.2 for CC jets with $E^{j}_{T} \approx 15 \; {\rm GeV}$.) Because there
is no electromagnetic coverage in the ICR, we do not apply a lower bound
cut on {\it emf}\hspace{0.5mm} in that region.

\begin{table}
\caption{Electron ID efficiencies used in this analysis.}
\begin{tabular}{ccc}
Detector Region & CC & EC \\ \hline
$\varepsilon^{e}_{\rm id}$ & $0.674 \pm 0.039$ & $0.242 \pm 0.075$ \\
\end{tabular}
\label{tbl:eid_eff}
\end{table}

A multijet data sample corrected for detector noise is used to measure
the jet identification efficiency, $\varepsilon^{j}_{\rm id}$. The
efficiency is a function of $E^{j}_{T}$, and is parametrized as in
Eq.~\ref{eqn:jid_eff_eqn}, with the fitted values of the parameters
listed in Table~\ref{tbl:jid_eff_param}.

\begin{equation}
\varepsilon^{j}_{\rm id} = p_{0} + p_{1}\times E^{j}_{T} + p_{2} \times (E^{j}_{T})^{2}.
\label{eqn:jid_eff_eqn}
\end{equation}

\begin{table*}
\caption{Parameters for jet identification efficiency as defined in 
Eq.~\ref{eqn:jid_eff_eqn}.}
\squeezetable
\begin{tabular}{ccccc}
Fiducial Region & $E^{j}_{T} \; (\rm{GeV})$ & $p_{0}$ & $p_{1} \; (\rm{GeV}^{-1})$ & $p_{2} \; (\rm{GeV}^{-2})$ \\
\tableline
CC & $15$--$27.4$ & $0.8994 \pm 0.0070$ & $(5.04 \pm 0.45) \times 10^{-3}$ & $(-6.7 \pm 1.0) \times 10^{-5}$ \\
($|\eta^{j}_{\rm d}| < 1.0$) & $\geq 27.4$ & $0.9864 \pm 0.0005$ & $(2.16 \pm 0.57) \times 10^{-5}$ & $(-1.90 \pm 0.30) \times 10^{-7}$ \\ 
\tableline
ICR & $15$--$30.5$ & $0.9838 \pm 0.0017$ & $(9.76 \pm 1.33) \times 10^{-4}$ & $(-1.76 \pm 0.27) \times 10^{-5} $ \\
($1.0 < |\eta^{j}_{\rm d}| < 1.5$) & $\geq 30.5$ & $0.9981 \pm 0.0008$ & $(-2.27 \pm 2.26) \times 10^{-5}$ & $(-1.52 \pm 1.22) \times 10^{-7}$ \\
\tableline
EC & & $0.9866 \pm 0.0004$ & $(-3.81 \pm 1.05) \times 10^{-5}$ & $(-1.15 \pm 0.75) \times 10^{-7} $ \\
($1.5 < |\eta^{j}_{\rm d}| < 2.5$) & & & &
\end{tabular}
\label{tbl:jid_eff_param}
\end{table*}

\subsubsection{Muons}
To avoid overlapping with the dilepton analysis, we veto events
containing isolated muons satisfying all the following criteria:

\begin{itemize}
\item The muon has a good track originating from the interaction vertex.
\item The muon has pseudorapidity $|\eta^{\mu}_{\rm d}| \leq 2.5$.
\item There is a large integrated magnetic field along the muon
trajectory ($\int \vec{B} \cdot d\vec{l} \;$). This ensures that the
muon traverses enough of the field to give a good $P_{T}$ measurement.
\item The energy deposited in the calorimeter along a muon track is at
least that expected from a minimum ionizing particle.
\item Transverse momentum $p_{T} \geq 4 \; \rm{GeV}$.
\item The distance in the $\eta - \phi$ plane between the muon and the
closest jet is $\Delta R(\mu,j) > 0.5$.
\end{itemize}

\subsubsection{Event selection} \label{subsubsec:selection}
About 1.9 million events passed the ELE\_JET\_HIGH and the
ELE\_JET\_HIGHA triggers. We require at least one electromagnetic
cluster with $E_{T} > 18 \; {\rm GeV}$ and a track matched to it. The
interaction vertex must be within $|z_{v}| < 60 \;
\rm{cm}$. About 600,000 events remain after these
selections. Kinematic and fiducial requirements are then applied to
select our base data sample. The criteria are listed below, with numbers
in the curly brackets specifying the number of events surviving the
corresponding requirement.

\begin{itemize}
\item One electron in the good fiducial volume ($|\eta^{e}_{\rm d}| < 1.1$ or
$1.5 < |\eta^{e}_{\rm d}| < 2.5$) passing restrictive electron
identification criteria, and with $E^{e}_{T} > 20 \;
\rm{GeV}$~---~\{15547\}.
\item No extra electrons in the good fiducial volume passing ``loose''
electron identification for $E^{e}_{T} > 15 \; \rm{GeV}$. The selection
criteria for the ``loose'' electrons are the same as those used for
signal electrons in the dilepton analysis, keeping two analyses
independent of each other~---~\{15319\}.
\item $|\eta^{e}| < 2.0$~---~\{13997\}.
\item No isolated muons~---~\{13980\}.
\item Four or more jets with $E_{T}^{j} > 15 \; \rm{GeV}$ and
$|\eta^{j}_{\rm d}| < 2.5$~---~\{187\}.
\item $\mbox{\MET} > 25 \; \rm{GeV}$~---~\{72\}.
\end{itemize}

After these selections the base sample contains 72 events. The major SM
backgrounds are from $W+ \geq 4$ jets $\rightarrow e + \nu + \geq 4$
jets, $t\overline{t} \rightarrow Wb\;Wb \rightarrow e + \nu + \geq 4$
jets, $WW + \geq 2 \; \mbox{jets} \rightarrow e + \nu + \geq 4$ jets,
and multijet events in which one of the jets is misidentified as an
electron and the jet transverse energies are inaccurately measured to
give rise to \MET.

\section{EVENT SIMULATION} \label{sec:fmc0}

We use {\sc pythia}~\cite{pythia} to simulate mSUGRA signal and
\ttbar\ and $WW$ backgrounds. We check our results
and obtain generator-dependent systematic errors using the {\sc
herwig}~\cite{herwig} generator. $W$~boson and associated jet production
is generated using {\sc vecbos}~\cite{vecbos} and {\sc herwig}. The
final state partons, which are generated by {\sc vecbos} as a result of
a leading order calculation, are passed through {\sc herwig} to include
the effects of additional radiation and the underlying processes, and to
model the hadronization of the final state partons~\cite{topmass_prd}.

\begin{figure*} \centering
        \epsfig{file=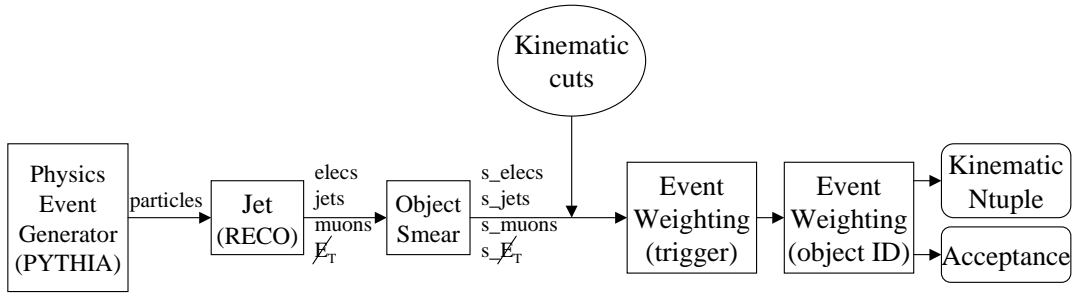, width=0.4\hsize, angle=270}
        \caption{Flow-chart of {\sc fmc\o}. Prefix ``s\_'' refers to
        smeared objects.}
	\label{fig:fmc0_flow}
\end{figure*}

In order to efficiently search for mSUGRA in a large parameter space and
to reduce the statistical error on signal acceptance, we used a fast
Monte Carlo program called {\sc fmc\o}~\cite{genik_thesis} to model
events in the D\O\ detector and to calculate the acceptance for any
physics process passing our trigger and offline selections. The
flow-chart of {\sc fmc\o} is shown in Fig.~\ref{fig:fmc0_flow}. First,
through a jet-reconstruction program, the stable particles that interact
in the detector are clustered into particle jets, in a way similar to
the clustering of calorimeter cells into jets. However, the generated
electrons, if they are not close to a jet ($\Delta R > 0.5$ in
$\eta-\phi$ space), are considered as the electrons reconstructed in the
detector. Otherwise, they are clustered into the jet. The generated
muons are considered as the reconstructed muons in the detector. Next,
the electrons, jets, muons, and \MET\ in the events are smeared
according to their resolutions determined from
data~\cite{topmass_prd}. The offline selections
(Sec.~\ref{subsubsec:selection}) are applied to the smeared
objects. Finally, each passed event is weighted with trigger and
identification efficiencies. The outputs of {\sc fmc\o} are an
``ntuple'' that contains the kinematic characteristics (\ET, $\eta$,
$\phi$, etc.) of every object and a run-summary ntuple that contains the
information of trigger efficiency and total acceptance for the process
being simulated. The acceptance $A$ is calculated as follows:

\begin{equation}
A = \frac{1}{N_{\rm gen}} \sum_{i}^{N_{\rm pass}} \varepsilon^{\rm total}_{\rm trig} \cdot \varepsilon^{e}_{\rm id} \cdot \varepsilon^{\rm jets}_{\rm id},
\label{eqn:acc_definition}
\end{equation}

\noindent where $\varepsilon^{\rm total}_{\rm trig}$ is the overall
trigger efficiency, $\varepsilon^{e}_{\rm id}$ is the electron
identification efficiency, $\varepsilon^{\rm jets}_{\rm id}$ is the
product of jet identification efficiencies of the four leading jets,
$N_{\rm gen}$ is the number of generated events, and $N_{\rm pass}$ is
the number of events that pass the offline kinematic requirements. The
uncertainty on the acceptance, $\delta_{A}$, is calculated as:

\begin{equation}
\delta_{A} = \frac{1}{N_{\rm gen}} \sum_{i}^{N_{\rm pass}} \delta_{\varepsilon},
\label{eqn:acc_err_definition}
\end{equation}

\noindent where $\delta_{\varepsilon}$ comes from the propagation of
uncertainties on $\varepsilon^{\rm total}_{\rm trig}$,
$\varepsilon^{e}_{\rm id}$, and $\varepsilon^{\rm jets}_{\rm id}$. Since
the same electron and jet identification efficiencies, and the same
trigger turn-ons are used the error on the acceptance is 100\%
correlated event-by-event as shown in Eq.~\ref{eqn:acc_err_definition}.

Because the signal triggers impose a combination of requirements on the
electron, jets, and \MET, the overall trigger efficiency has three
corresponding components. The efficiency of each component was measured
using data. The individual efficiencies are then used to construct the
overall trigger efficiency. The details of the measurements and
construction are documented in
Ref.~\cite{zhou_thesis}. Table~\ref{tbl:W+jets_trig_eff} compares the
trigger efficiencies of $W +$~jets events measured in data with those
simulated using {\sc vecbos} Monte Carlo. We find that they are in good
agreement at each jet multiplicity.

\begin{table}
\caption{Comparison of $\varepsilon^{total}_{trig}$, the total trigger
efficiency of ELE\_JET\_HIGH trigger. The second column lists the
efficiencies measured using $W +$~jets data; the third column lists the
simulated efficiencies found by putting the {\sc vecbos} $W+$~jets
events through {\sc fmc\o}.}
\begin{tabular}{ccc}
$N_{\rm jet}$ & Data & {\sc vecbos} \\
\tableline
$\geq 1$ & $0.589 \pm 0.019$ & $0.579 \pm 0.022$ \\
$\geq 2$ & $0.826 \pm 0.027$ & $0.833 \pm 0.020$ \\
$\geq 3$ & $0.928 \pm 0.031$ & $0.925 \pm 0.016$ \\
$\geq 4$ & $0.944 \pm 0.037$ & $0.957 \pm 0.012$ \\
\end{tabular}
\label{tbl:W+jets_trig_eff}
\end{table}

We also compared the acceptance of {\sc fmc\o} with {\sc
geant}~\cite{geant} and data, and found good agreement for $W +$~jets,
\ttbar, and $WW$ events.

\section{BACKGROUNDS} \label{sec:backgrounds}

\subsection{Multijet background} \label{subsec:qcd_bkgd}
From the ELE\_JET\_HIGH and ELE\_JET\_HIGHA triggered data we obtain two
sub-samples. For sample~1, we require all offline criteria to be
satisfied, except for \MET. At small \MET\ ($< 20 \; {\rm GeV}$),
sample~1 contains contributions mainly from multijet production, where
jet energy fluctuations give rise to \MET. At large \MET\ ($> 25 \; {\rm
GeV}$), it has significant contributions from $W+$~jets events, with
additional contributions from \ttbar\ production and possibly the mSUGRA
signal. For sample~2, we require that the EM object represent a very
unlikely electron candidate by applying an ``anti-electron''
requirement~\cite{zhou_thesis}. All other event characteristics are the
same as those in sample~1. The sample~2 requirements tend to select
events in which a jet mimics an electron, and consequently sample~2
contains mainly multijet events with little contribution from other
sources for $\mbox{\MET} > 25 \; {\rm GeV}$. The \MET\ spectra of the
two samples can therefore be used to estimate the number of multijet
background events (\Nqcd) in sample~1 as follows. We first normalize the
\MET\ spectrum of sample~2 to that of sample~1 in the low-\MET\ region,
and then estimate \Nqcd\ by multiplying the number of events in the
signal region ($\mbox{\MET} > 25 \; \rm{GeV}$) of sample~2 by the same
relative normalization factor~\cite{yu_thesis}.

\begin{figure}
	\epsfig{file=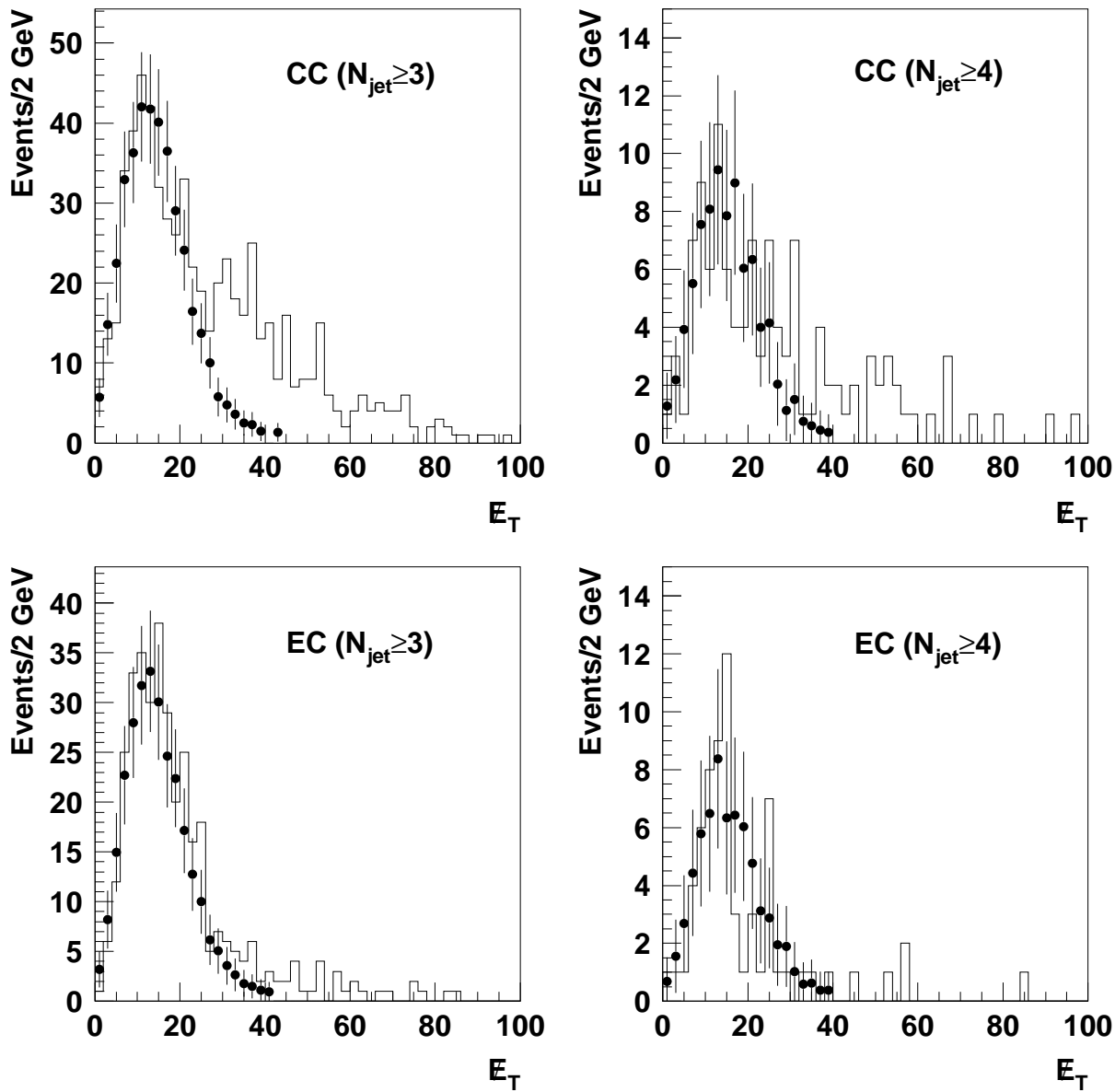,width=\hsize}
	\caption{\MET\ spectra of sample~2 (points) normalized to
	sample~1 (histograms) in the region of $0 \leq \mbox{\MET} \leq
	14 \; {\rm GeV}$. The normalizations are done for the fake
	electron in the CC and EC, respectively. The errors are
	statistical only.}
	\label{fig:ejh_qcd_bkgd}
\end{figure}

The \MET\ spectra for both samples are shown in
Fig.~\ref{fig:ejh_qcd_bkgd}, normalized to each other for $0 \leq
\mbox{\MET} \leq 14 \; {\rm GeV}$, and for the cases in which the fake
electron is in the CC and EC, respectively. From these distributions, we
calculate \Nqcd\ to be $82.6 \pm 15.3$ and $19.1 \pm 4.7$, for inclusive
jet multiplicities of 3 and 4 jets, respectively. (The inclusive 3-jet
sample is obtained the same way as the base sample, except that we
require at least 3 jets, rather than 4, in the event.) The errors
include statistical uncertainties and systematic uncertainties in the
trigger and object identification efficiencies, different definitions of
sample~2, and different choice for the normalization regions.

\subsection{$t\overline{t}$ background} \label{subsec:ttbar_bkgd}
The number of \ttbar\ background events, \Ntt, is calculated using {\sc
fmc\o}. The \ttbar\ events were generated using {\sc
pythia}~\cite{pythia} for $m_{\rm top}=175 \; \rm{GeV}$. A \ttbar\
production cross section of $\sigma = 5.9 \pm 1.7 \; \rm{pb}$, as
measured by D\O~\cite{ttbar_xsec_prd}, is used. The results are
$\mbox{\Ntt}=27.7 \pm 8.3$ events and $\mbox{\Ntt}=16.8 \pm 5.2$ events
for inclusive jet multiplicities of 3 and 4 jets, respectively. The
errors include uncertainties on the \ttbar\ production cross section,
differences in physics generators, trigger and object identification
efficiencies, and on the integrated luminosity.

\subsection{$WW+$ jets background} \label{subsec:ww_bkgd}
{\sc fmc\o} is also used to calculate the $WW+$~jets background. The
production cross section at next-to-leading order is taken as $\sigma =
10.40 \pm 0.23 \; \rm{pb}$~\cite{ww_xsec,single_top}, assuming no
anomalous couplings ($\delta\kappa = \lambda = 0$)~\cite{ww_anomaly}. The
$WW$ events were generated using {\sc pythia}. There are $7.7 \pm 1.2$
and $1.4 \pm 0.3$ events expected for inclusive jet multiplicity of 3
and 4 jets, respectively. The errors include uncertainties on the
production cross section, trigger and object identification
efficiencies, differences in physics generators, the jet energy scale,
and on the integrated luminosity.

\subsection{$W+$ jets background} \label{subsec:wjets_bkgd}
To good approximation, each extra jet in $W+$~jets events is the result
of an extra coupling of strength $\alpha_{s}$~\cite{vecbos}, and we
expect the number of $W+$~jets events to scale as a power of $N_{\rm
jet}$. The scaling law is supported by the $W+$~jets, $Z+$~jets, and
$\gamma+$~jets data~\cite{ttbar_xsec_prl}. In this analysis, we first
estimate the number of $W+ \geq 3$-jet events, $N^{W}_{3}$, in the data
collected with ELE\_JET\_HIGH and ELE\_JET\_HIGHA triggers, and then
extract the effective scaling factor $\alpha$ using $W+ \geq n$-jet
events collected with EM1\_EISTRKCC\_MS trigger. The expected number of
$W+ \geq 4$-jet events ($N^{W}_{4}$) in our base sample is then:

\begin{equation}
N^{W}_{4} = N^{W}_{3} \cdot \alpha \cdot \frac{\varepsilon^{W4}_{\rm trig}}{\varepsilon^{W3}_{\rm trig}},
\label{eqn:scaling_law_1}
\end{equation}

\noindent where $\varepsilon^{W3}_{\rm trig}$ and $\varepsilon^{W4}_{\rm 
trig}$ are trigger efficiencies of $W+ \geq 3$-jet and $W+ \geq 4$-jet
events, respectively, as shown in Table~\ref{tbl:W+jets_trig_eff}.

\subsubsection{Estimating the number of $W+ \geq n$-jet events}
\label{subsubsec:nw}
We estimate the number of $W+ \geq n$-jet events in a way similar to
that used to estimate the multijet background. We first use a neural
network (NN) to define a kinematic region in which $W+\geq n$-jet events
dominate the background and any possible contribution from mSUGRA can be
neglected. In that region, we normalize the number of $W+ \geq n$-jet MC
events to the number of events observed in the data which have had all
other major SM backgrounds subtracted. The normalization factor is then
applied to the whole $W+ \geq n$-jet MC sample to obtain our estimate
for the $W+ \geq n$-jet background in the data.

In this analysis, we use a NN package called {\sc
mlpfit}~\cite{mlpfit}. All NNs have the structure of X-2X-1, where X is
the number of input nodes, i.e., the number of variables used for
training, and 2X is the number of nodes in the hidden layer. We always
use 1 output node with an output range of 0 to 1. Signal events (in this
case, $W+ \geq n$-jet events) are expected to have NN output near 1 and
background events near 0. We choose the NN output region of 0.5--1.0 to
be the ``signal''-dominant kinematic region. The variables used to
distinguish $W+ \geq n$-jet events from other SM backgrounds and the
mSUGRA signal are:

\begin{itemize} \label{nn_var1}
\item \MET\
\item $E^{e}_{T}$
\item $H_{T} = \sum E^{j}_{T}$ for all jets with $E^{j}_{T}>15 \; {\rm GeV}$
\item $\Delta \phi_{e,\mbox{\scriptsize \MET}}$
\item $M_{T} = \sqrt{2 E^{e}_{T} \mbox{\MET}[1-\cos(\Delta\phi_{e,\mbox{\scriptsize \MET}})]}$
\item $\Delta \phi_{j_{1},\mbox{\scriptsize \MET}}$ (not used for $\geq 4$-jet events)
\item $\Delta \phi_{j_{2},\mbox{\scriptsize \MET}}$ (used for $\geq 2$-jet and
$\geq 3$-jet events)
\item $\mathcal{A}$---aplanarity~\cite{collider_physics} (used for $\geq
2$, $\geq 3$, and $\geq 4$-jet events) is defined in terms of the
normalized momentum tensor of the $W$ boson and the jets with
$E^{j}_{T}>15 \; \rm{GeV}$:
\begin{equation}
M_{ab} = \frac{\sum_{i} p_{ia}p_{ib}}{\sum_{i} p^{2}_{i}},
\end{equation}
\noindent where $\vec{p_{i}}$ is the three-momentum of object $i$ in the
laboratory frame, and $a$ and $b$ run over the $x$, $y$, and $z$
coordinates. Denoting $Q_{1}$, $Q_{2}$, and $Q_{3}$ as the three
eigenvalues of $M_{ab}$ in ascending order, $\mathcal{A}$ $= 1.5
\times Q_{1}$. The $p_{z}$ of the $W$ boson is calculated by imposing
the requirement that the invariant mass of the electron and the neutrino
(assumed to be the source of \MET) equals the $W$ boson mass. This
requirement results in a quadratic equation for the longitudinal
momentum of the neutrino. Because the probability of a small $p_{z}$ is
usually higher than that of a large $p_{z}$, the smaller $p_{z}$
solution is always chosen. In cases where there is no real solution,
\MET\ is increased until a real solution is obtained.
\item $r_{H} = H_{T2}/H_{Z}$, where $H_{T2} = H_{T}-E^{j_{1}}_{T}$, and
$H_{Z} = \sum_{i} |p_{z}|$ where $i$ runs over the electron, all jets
with $E^{j}_{T} > 15 \; {\rm GeV}$, and neutrino (as assumed in the
calculation of $\mathcal{A}$) in the event~\cite{top_mass_prd} (only
used for $\geq 4$-jet events).
\item $\cos\theta^{*}_{e}$, where $\theta^{*}_{e}$ is the polar
angle of the electron in the $W$ boson rest frame, relative to the
direction of flight of the $W$ boson. The $W$ boson four-momentum is
obtained by fitting the event to a \ttbar\ assumption. The details of
the fit are described in Ref.~\cite{top_mass_prd} (only used for $\geq
4$-jet events).
\item $\cos{\theta}^{*}_{eb}$, where $\theta^{*}_{eb}$ is the angle
between the electron and the $b$ jet from the same top (or antitop)
quark in the $W$ boson rest frame~\cite{top_spin}. Again, a fit to the
\ttbar\ assumption is performed to identify the correct $b$ jet (only
used for $\geq 4$-jet events).
\end{itemize}

All the offline requirements described in Sec.~\ref{subsubsec:selection}
are applied except that the requirement on the number of jets is reduced
corresponding to different inclusive jet multiplicity. The multijet,
\ttbar, and $WW$ backgrounds are estimated using the methods described
in Sec.~\ref{subsec:qcd_bkgd}--\ref{subsec:wjets_bkgd}. The mSUGRA events
were generated with $m_{0}=170 \; {\rm GeV}$, $m_{1/2}=58 \; {\rm GeV}$
and $\tan\beta = 3$. This parameter set was chosen because it is close
to the search limit obtained in the dilepton analysis.

\subsubsection{Estimating $N^{W}_{3}$} \label{subsubsec:nw3}
\begin{figure}
\begin{minipage}{0.5\linewidth}
  \centering\epsfig{file=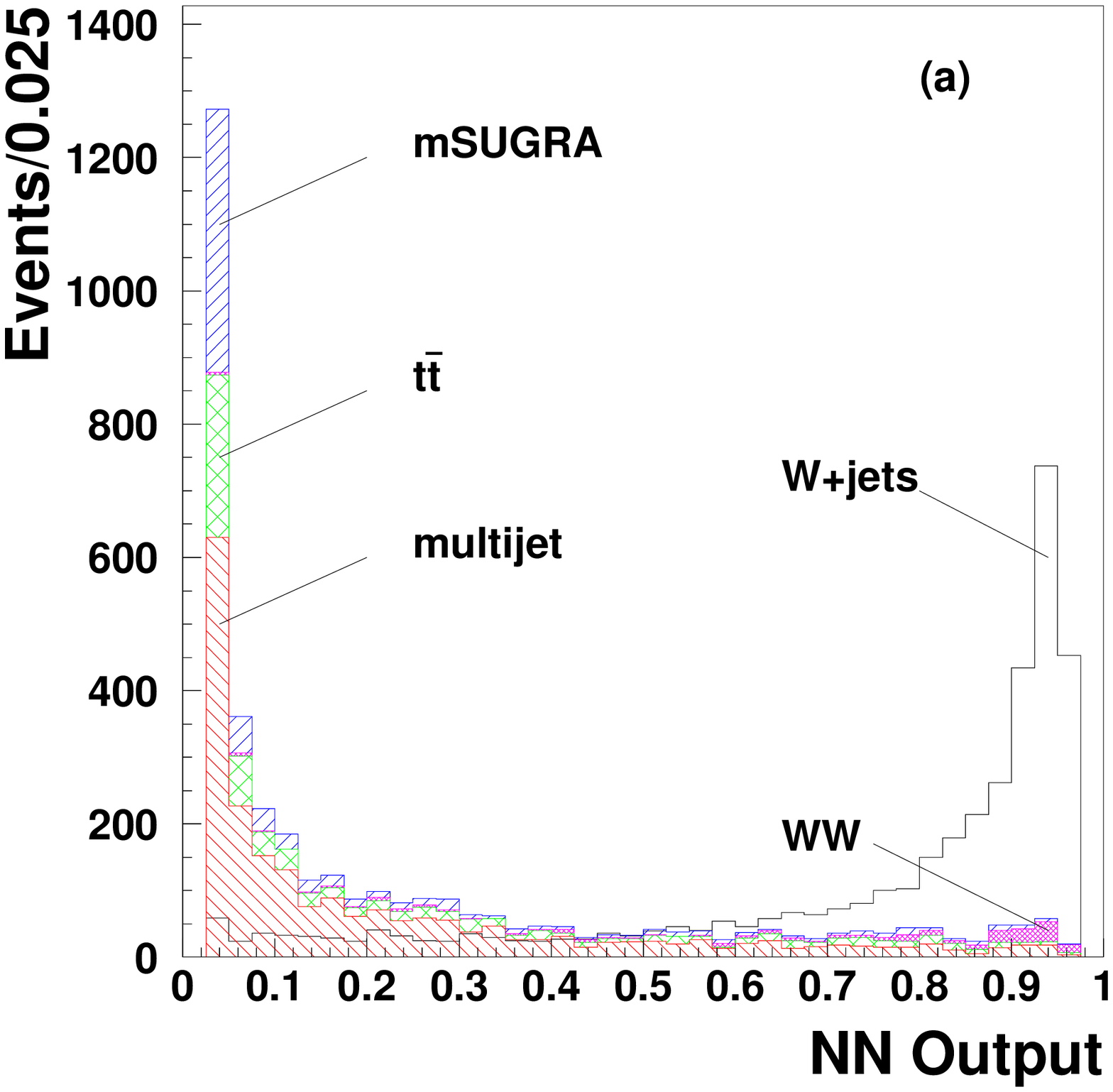,width=\linewidth}
\end{minipage}\hfill
\begin{minipage}{0.5\linewidth}
  \centering\epsfig{file=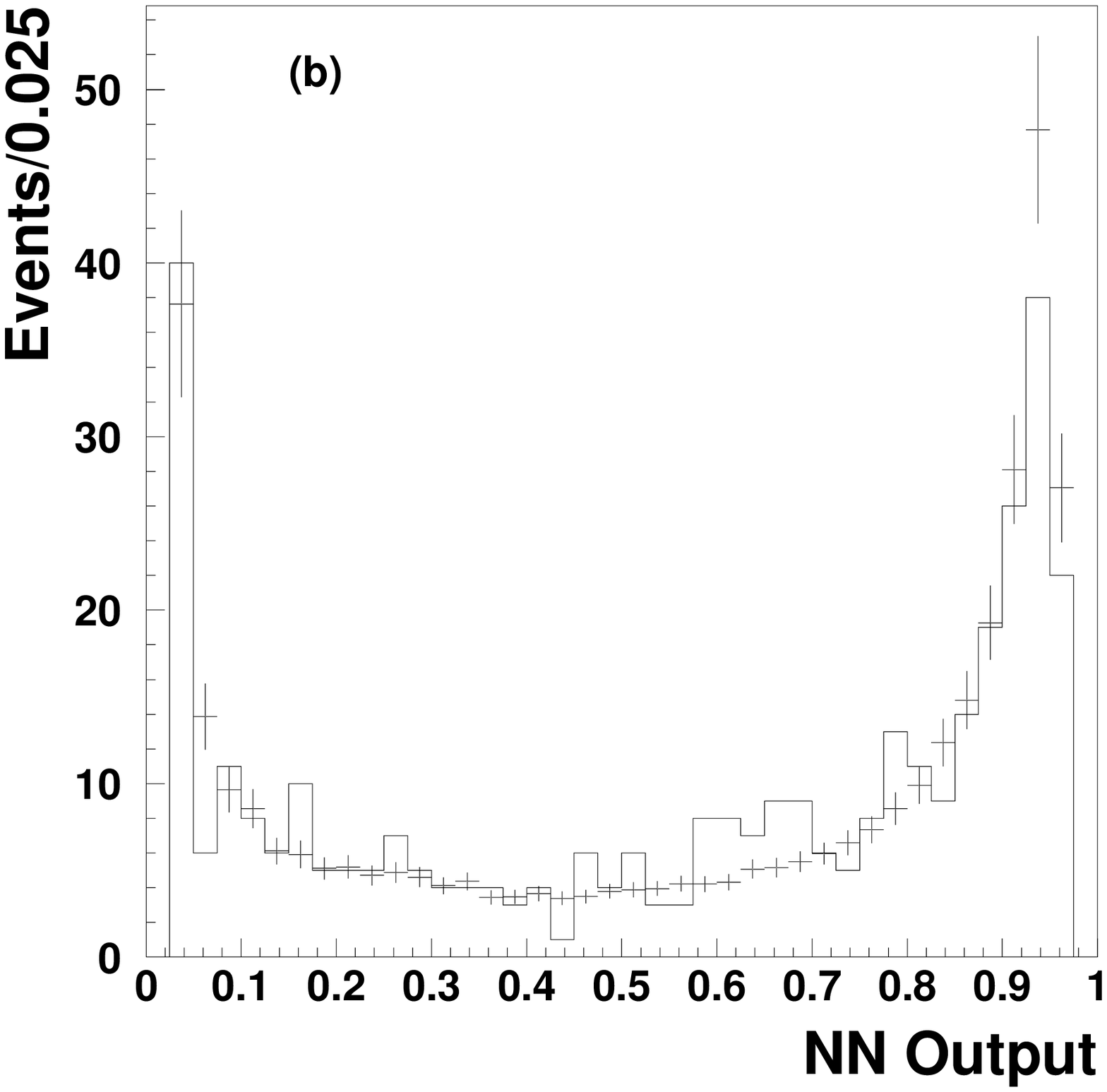,width=\linewidth}
\end{minipage}
  \caption{(a) Expected NN output for events passing the ELE\_JET\_HIGH
  or ELE\_JET\_HIGHA triggers and with $\geq 3$~jets. (b) Expected NN
  output for data (points) and the observed NN output for data
  (histogram). The error on the points include statistical and
  systematic errors. All events were required to pass our offline
  selections, except that we required only 3~jets instead of 4.}
  \label{fig:ejh_nn_3jets_dphi_ej}
\end{figure}

The result of the NN training for $\geq 3$-jet events is shown in
Fig.~\ref{fig:ejh_nn_3jets_dphi_ej}(a). The number of $W + \geq 3$-jet
events used in the training is the same as the sum of all background
events, including any possible mSUGRA sources in their expected
proportions. The match between training and data is shown in
Fig.~\ref{fig:ejh_nn_3jets_dphi_ej}(b), where the data and MC are
normalized to each other for NN output between 0.5 and 1.0. Because the
number of mSUGRA events is negligible in this region, we do not include
them in the background subtraction. We estimate that $241.8 \pm 18.0$ $W
+ \geq 3$-jet events pass our final 3-jet selection.

\subsubsection{Measuring the scaling factor $\alpha$}
\label{subsubsec:alpha_susy}
We extract the parameter $\alpha$ from the data passing the
EM1\_EISTRKCC\_MS trigger, which does not have a jet requirement in the
trigger, and fit the measured number of $W+n$-jet events
($\overline{N}^{W}_{n}$) to:

\begin{equation}
\overline{N}^{W}_{n} = \overline{N}^{W}_{1} \cdot \alpha^{n-1}.
\label{eqn:scaling_law_2}
\end{equation}

$\overline{N}^{W}_{n}$ values are obtained as described in
Sec.~\ref{subsubsec:nw}. The NN training and normalization to the data
are performed separately for each inclusive jet multiplicity. The
results are summarized in Table~\ref{tbl:eis_wjets}. The errors on
$\overline{N}^{W}_{n}$ include statistical errors from MC and data, and
uncertainties on the choice of different normalization regions and on
the choice of different QCD dynamic scales used in generating {\sc
vecbos} events.

\begin{figure} \centering
        \epsfig{file=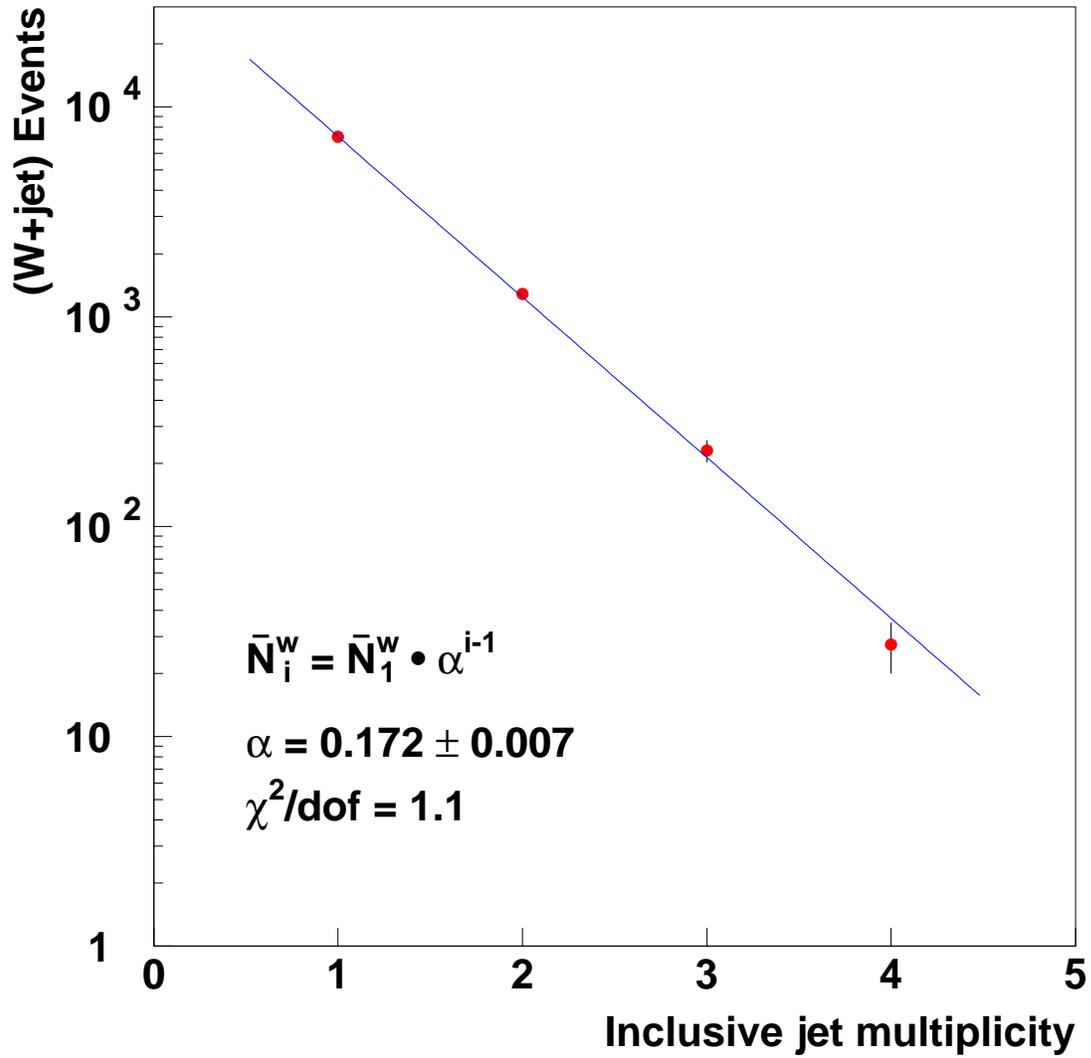, width=0.9\hsize}
        \caption{Fit of $W+ \geq n$-jet events to the power law of
        Eq.~\ref{eqn:scaling_law_2}.}
	\label{fig:fit_wjets_susy}
\end{figure}

The fit of $\overline{N}^{W}_{n}$ to Eq.~\ref{eqn:scaling_law_2} is
shown in Fig.~\ref{fig:fit_wjets_susy}, from which we extract $\alpha =
0.172 \pm 0.007$.

\vspace{1cm}

\small
\begin{table}
\caption{Estimated number of $W+ \geq n$-jet events,
$\overline{N}^{W}_{n}$, as a function of inclusive jet multiplicity in
the data passing the EM1\_EISTRKCC\_MS trigger. They were obtained by
normalizing MC to data in the NN output region where $W+ \geq n$-jets
events dominate (see text). $\overline{N}_{\rm data}$ is the number of
observed events. The mSUGRA events were generated with $m_{0}=170 \;
{\rm GeV}$, $m_{1/2}=58 \; {\rm GeV}$, and $\tan\beta = 3$.}
\begin{tabular}{lcccc}
$N_{\rm jet}$ & $\geq 1$ & $\geq 2$ & $\geq 3$ & $\geq 4$ \\
\tableline \\[-2mm]
$\overline{N}_{\rm data}$ & $8191$ & $1691$ & $353$ & $64$ \\
$N_{\rm multijet}$ & $826 \pm 95$ & $291 \pm 48$ & $75 \pm 15$ & $16.6 \pm  7.0$ \\
$N_{t\overline{t}}$ & $25.8 \pm 7.6$ & $26.1 \pm 7.6$ & $21.9 \pm 6.5$ & $13.5 \pm 4.3$ \\
$N_{WW}$ & $33.7 \pm 3.3$ & $23.6 \pm 2.3$ & $6.19 \pm 0.95$ & $1.12 \pm 0.25$ \\
$\overline{N}^{W}_{n}$ & $7210 \pm 131$ & $1283 \pm 79$ & $230 \pm 27$ & $27.4 \pm 7.4$ \\
$N_{\rm mSUGRA}$ & $28.3 \pm 3.7$ & $25.0 \pm 3.1$ & $19.7 \pm 2.7$ & $12.6 \pm 2.1$ \\
\end{tabular}
\label{tbl:eis_wjets}
\end{table}
\normalsize

\subsubsection{Calculating the number of $W+ \geq 4$-jet events,
$N^{W}_{4}$} \label{subsec:nw4}

With $\varepsilon^{W3}_{\rm trig}=0.925 \pm 0.016$ and
$\varepsilon^{W4}_{\rm trig}=0.957 \pm 0.012$, and using
Eq.~\ref{eqn:scaling_law_1}, we obtain $N^{W}_{4} = 43.0 \pm 7.6$.

\subsection{Summary} \label{subsec:total_bkgd}
The expected numbers of events in the base data sample from the major
sources of background are summarized in
Table~\ref{tbl:total_background}. From the table, we conclude that the
sum of the backgrounds is consistent with the observed number of
candidate events.

\begin{table}
\caption{Expected numbers of events in the base data sample from the
major sources of background and the number of observed data events.}
\begin{tabular}{lc}
$W+\geq 4$-jets   & $43.0$ $\pm$ $7.6$  \\
misidentified multijet          & $19.1$ $\pm$ $4.7$  \\
\ttbar\           & $16.8$ $\pm$ $5.2$  \\
$WW+\geq 2$-jets  & $1.4$ $\pm$ $0.3$ \\[1mm]
\tableline
Total             & $80.3$ $\pm$ $10.4$ \\
Data              & 72 \\
\end{tabular}
\label{tbl:total_background}
\end{table}

\section{SEARCH FOR SIGNAL}

\subsection{Neural Network Analysis}
We use a NN analysis to define a kinematic region in which the
sensitivity of signal to background is highest. We use the following
variables in the NN. Those not defined below have been defined in
Sec.\ref{subsubsec:nw}.

\begin{itemize} \label{nn_var2}
\item \MET\ -- For the signal, \MET\ comes from two LSPs and at least
one neutrino. For the \ttbar, $W+$~jets, and $WW$ backgrounds, it comes
from the neutrino. For multijet background, it comes from fluctuation in
the measurement of the jet energy. Generally, the signal has larger \MET\
than the backgrounds.
\item $E^{e}_{T}$ -- The electron in the signal comes from a virtual $W$
boson decay. Its spectrum is softer than that of the electrons from the
\ttbar\ and $W+$~jets backgrounds.
\item \HT\ -- A pair of heavy mSUGRA particles are produced in the hard
scattering and most of the transverse energy is carried away by
jets. The $H_{T}$ for the signal thus tends to be larger than that for
the major backgrounds.
\item $E^{j_{3}}_{T}$ -- The third leading jet in $E_{T}$ from $W+$~jets,
$WW$, and multijet events most likely originates from gluon
emission. For \ttbar\ and mSUGRA events, it is probably due to $W$ boson
decay. Thus, the \ttbar\ and mSUGRA signals have a harder
$E^{j_{3}}_{T}$ spectrum.
\item \MT\ -- For \ttbar, $W+$~jet, and $WW$ events, \MT\ peaks near
$M_{W}=80 \; \rm{GeV}$. This is not the case for the signal since we
expect the $W$~boson produced in the decay chain to be virtual for a
wide range of $m_{1/2}$ up to 200 {\rm GeV}.
\item $\Delta \phi_{e,\mbox{\scriptsize \MET}}$ -- Because the electron and
neutrino form a $W$ boson in \ttbar, $W+$~jet, and $WW$ events, their
$\Delta \phi_{e,\mbox{\scriptsize \MET}}$ spectra should peak away from
$\Delta \phi_{e,\mbox{\scriptsize \MET}}=0$. For multijet events, the
$\Delta \phi_{e,\mbox{\scriptsize \MET}}$ spectrum should peak near 0
and $\pi$ because \MET\ can be caused by fluctuations in the energy of
the jet which mimics an electron.
\item $\mathcal{A}$ -- $W+$~jets, $WW$, and multijet events are more likely 
to be collinear due to QCD bremsstrahlung, while the signal and \ttbar\
events are more likely to be spherical.
\item $\cos\theta^{*}_{j}$, where $\theta^{*}_{j}$ is the polar angle
of the higher-energy jet from $W$ boson decay in the rest frame of
parent $W$ boson, relative to the direction of flight of the $W$
boson. This is calculated by fitting all the events to the \ttbar\
assumption. For \ttbar\ production, the spectrum is isotropic, but for
the signal and other SM backgrounds, it is not.
\item $\cos\theta^{*}_{e}$, the signal has a somewhat different
$\cos\theta^{*}_{e}$ distribution than the background does, especially
for \ttbar\ events.
\end{itemize}

\begin{figure}
        \epsfig{file=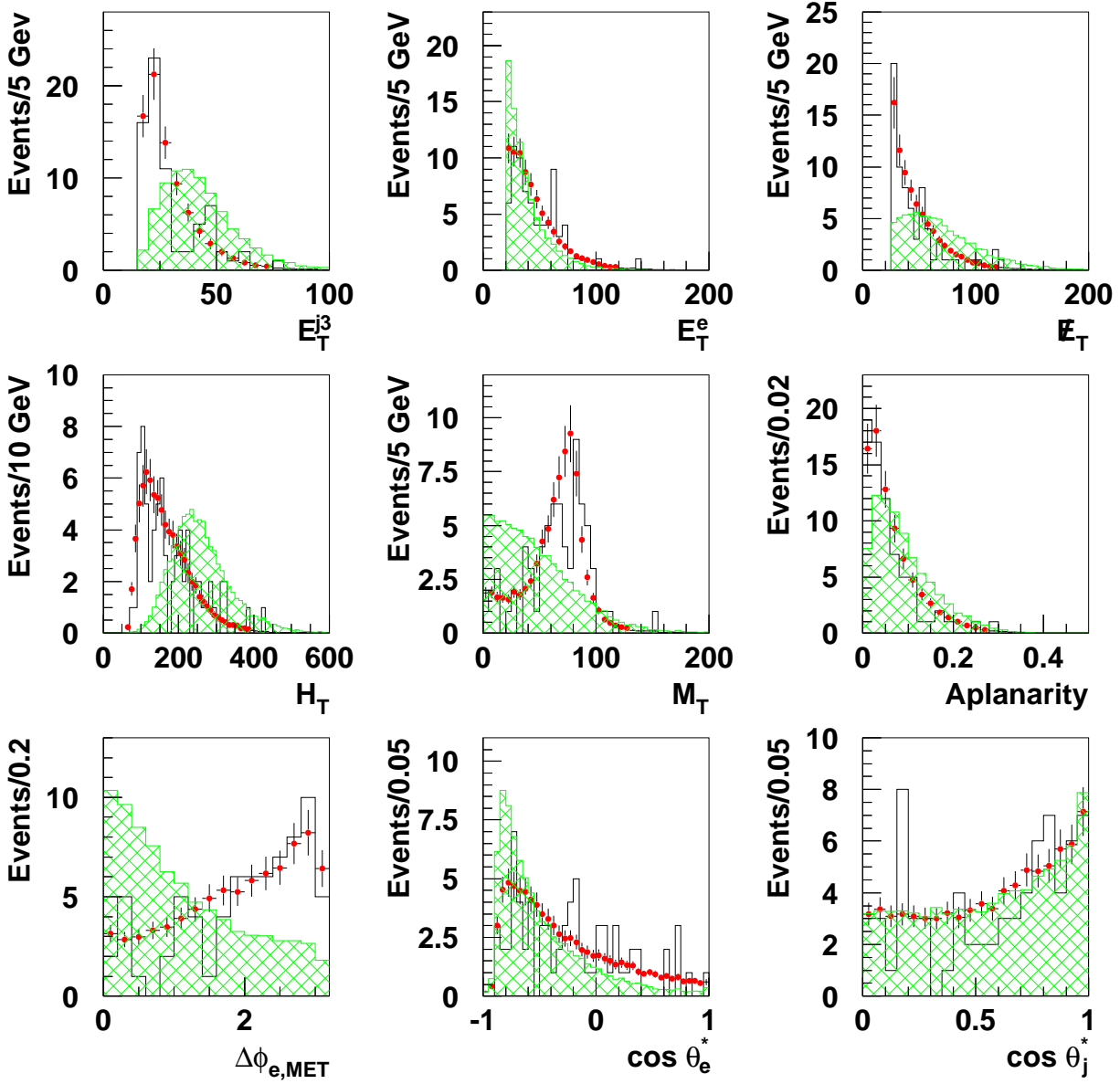, width=\hsize}
        \caption{Distribution of NN variables for data (open histogram),
        background (points) and signal (hatched histogram). The signal
        was generated at $m_{0}=170 \; \rm{GeV}$, $m_{1/2}=58 \;
        \rm{GeV}$, and $\tan\beta = 3$. We have multiplied the expected
        number of signal events (18.5) by a factor of 4.3 to normalize
        it to the total number of background events. Since the same
        number of signal and background events are used to train the NN,
        the plot shows the relative strength of signal to background as
        seen by the NN.}
        \label{fig:nn_var_susy}
\end{figure}

\begin{figure}
\begin{minipage}{0.5\linewidth}
  \centering\epsfig{file=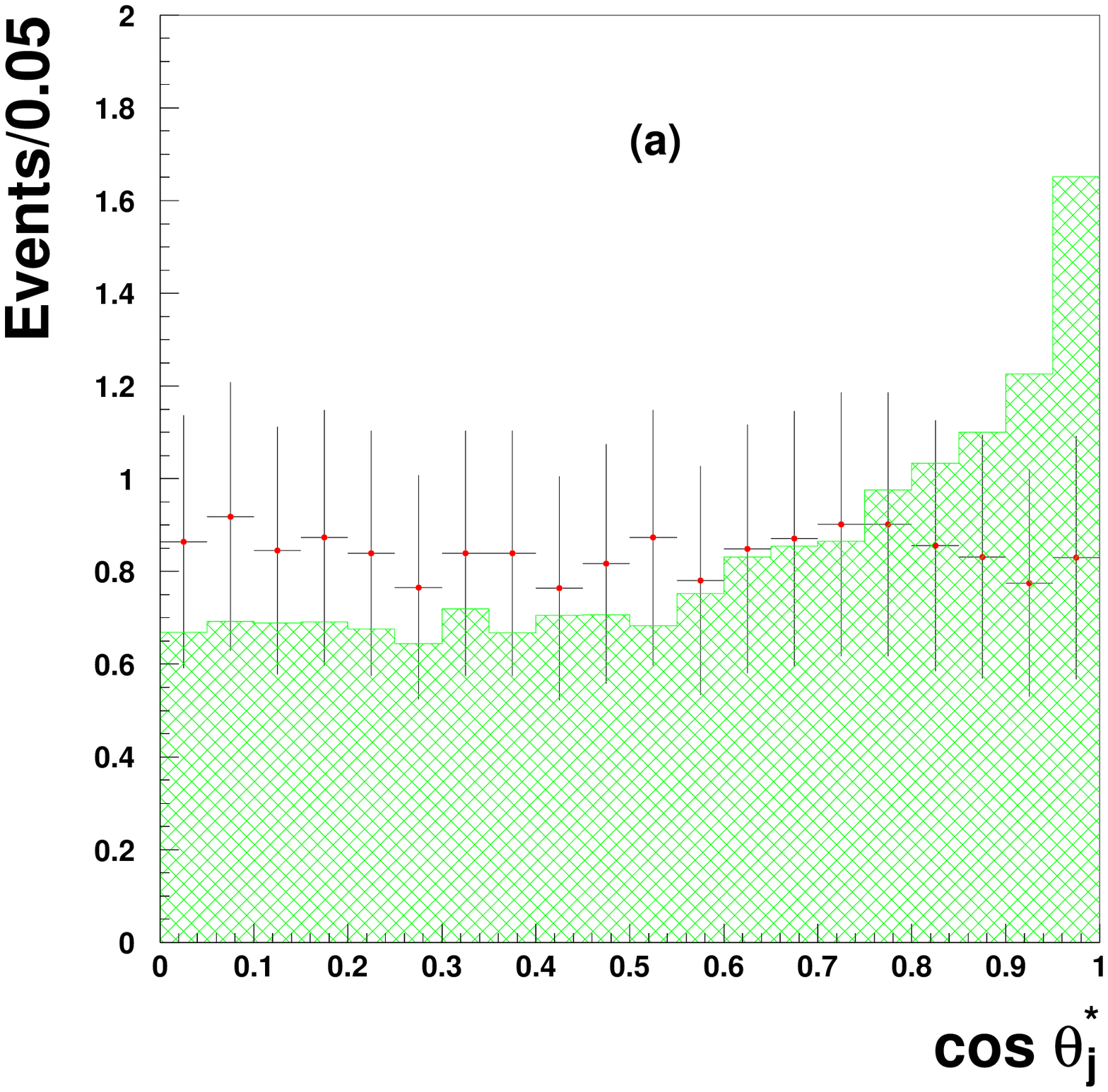,width=\linewidth}
\end{minipage}\hfill
\begin{minipage}{0.5\linewidth}
  \centering\epsfig{file=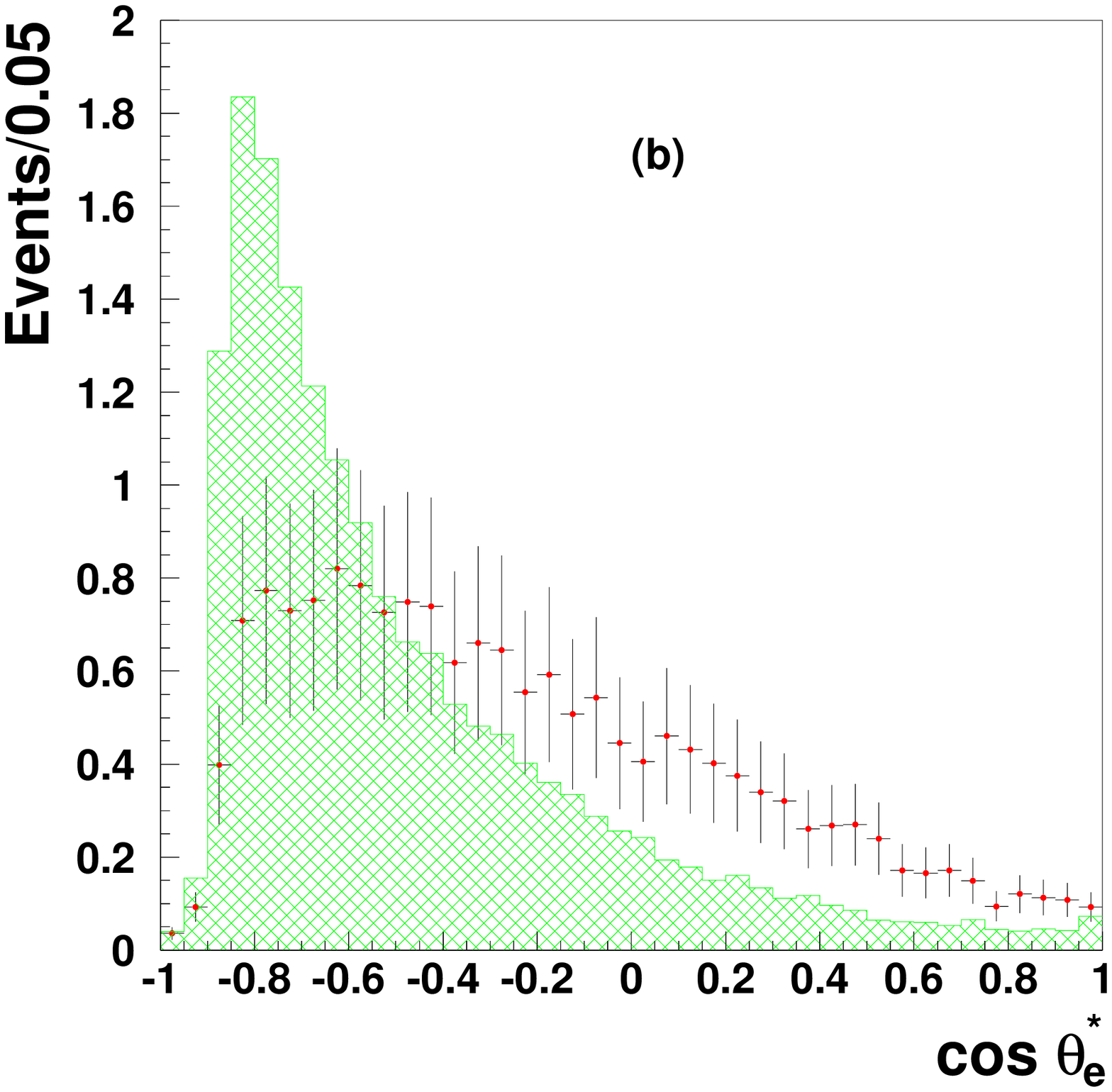,width=\linewidth}
\end{minipage}
  \caption{Distribution of (a) $\cos\theta^{*}_{j}$ and (b)
  $\cos\theta^{*}_{e}$ for signal (hatched histogram) and \ttbar\
  events (points). The signal was generated at $m_{0}=170 \;
  \rm{GeV}$, $m_{1/2}=58 \; \rm{GeV}$, and $\tan\beta = 3$. We have
  multiplied the expected number of signal events (18.5) by a factor of
  0.91 to normalize it to the number of \ttbar\ events expected in our
  base sample.}
\label{fig:sig_top_costh}
\end{figure}

\begin{figure}
        \epsfig{file=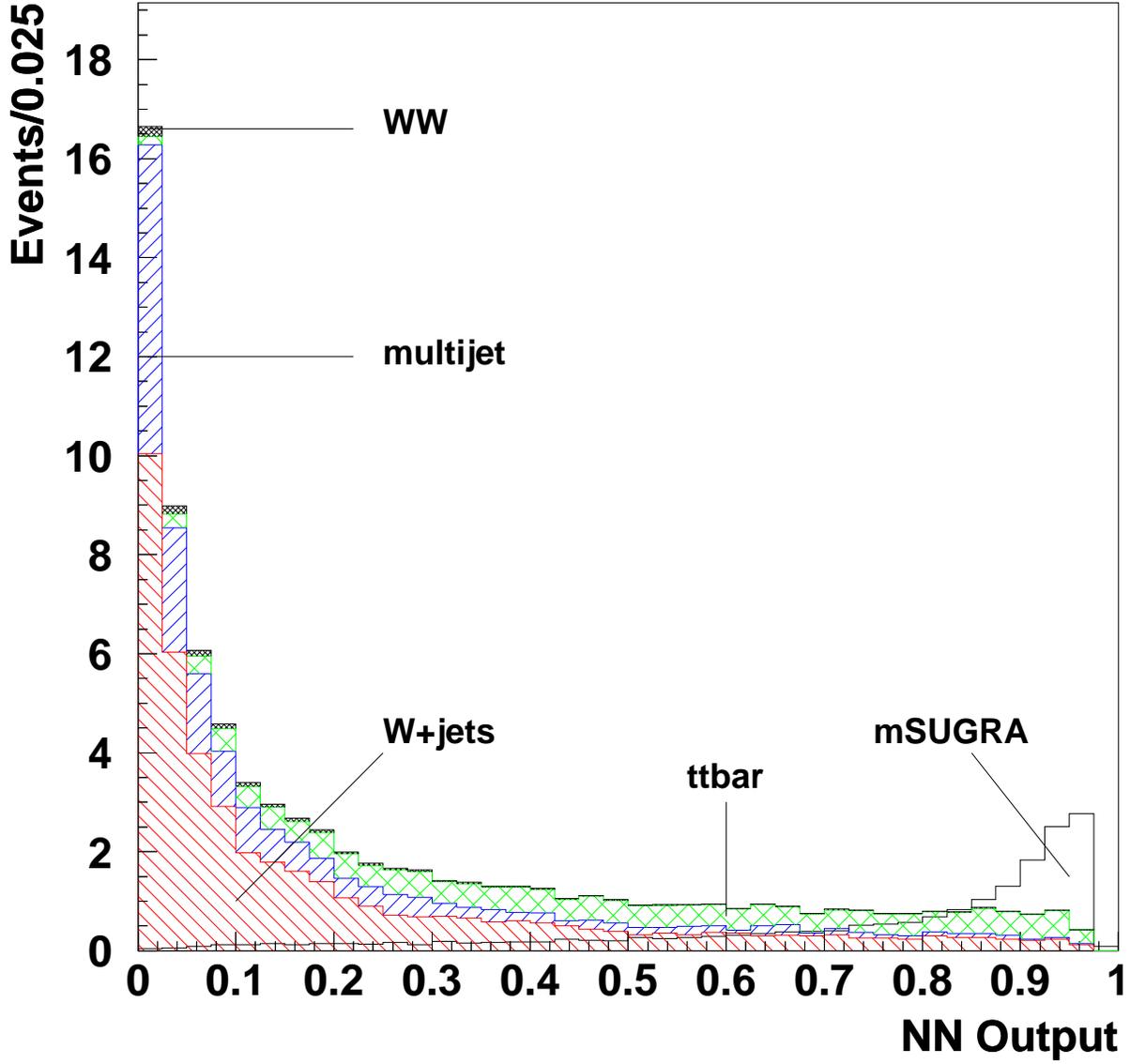, width=\hsize}
        \caption{Result of training of a NN. The excess above the
        background near 1 is the expected signal. The signal was
        generated at $m_{0}=170 \; \rm{GeV}$, $m_{1/2}=58 \; \rm{GeV}$,
        and $\tan\beta = 3$. The backgrounds are stacked up in the order
        of $W(e\nu)+$~jets, $W(\tau\nu)+$~jets, misidentified multijet,
        \ttbar, and $WW$ production. The contribution of each type of
        background is normalized to its expected number of events in the
        data.}
        \label{fig:nn_susy_bkgd}
\end{figure}

\begin{figure}
        \epsfig{file=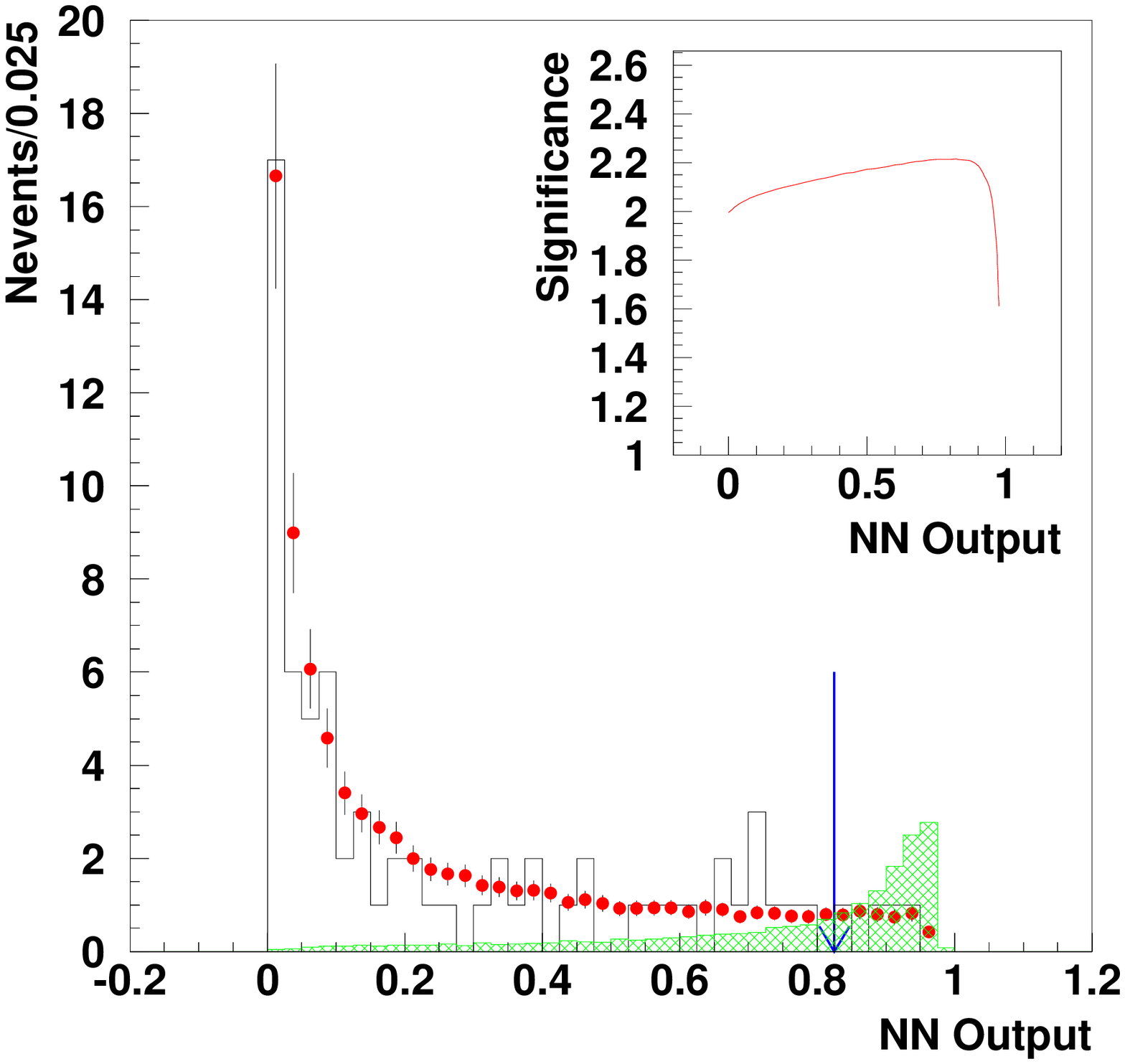, width=\hsize}
        \caption{NN output for data (open histogram), signal (hatched
        histogram), and background (points). The signal was generated at
        $m_{0}=170 \; \rm{GeV}$, $m_{1/2}=58 \; \rm{GeV}$, and $\tan\beta
        = 3$. The background expectation describes the data well. The
        vertical arrow indicates the cutoff on the NN output that
        corresponds to the maximum signal significance. The significance
        (described in Sec.~\ref{subsec:signif}) as a function of NN
        output is plotted in the insert.}
        \label{fig:nn_data_susy_bkgd}
\end{figure}

\noindent The spectra for these variables are shown in
Fig.~\ref{fig:nn_var_susy}. There is no evidence of an excess in our
data for the mSUGRA parameters used. Fig.~\ref{fig:sig_top_costh}
displays the $\cos\theta^{*}_{j}$ and $\cos\theta^{*}_{e}$
distributions for signal and \ttbar\ events. These two variables are
particularly useful in reducing the \ttbar\ background relative to the
mSUGRA signal. Nevertheless, \ttbar\ events still make the largest
contribution in the signal-rich region because of their similarity to
the mSUGRA signal. This can be seen in Fig.~\ref{fig:nn_susy_bkgd}, in
which the NN output is displayed for each background and the mSUGRA
signal for a particular set of parameters. The result of the NN output
for data is given in Fig.~\ref{fig:nn_data_susy_bkgd}. The expected
background describes the data well.

\subsection{Signal Significance} \label{subsec:signif}
To apply the optimal cut on the NN output, we calculated the signal
significance based on the expected number of signal ($s$) and background
($b$) events that would survive any NN cutoff. We define the
significance ($\overline{{\mathcal S}}$) below. The probability that the
number of background events, $b$, fluctuates to $n$ or more events is:

\begin{equation}
F(n|b) = \sum_{k=n}^{\infty} p(k|b) =
\frac{1}{\sqrt{2\pi}} \int_{{\mathcal S}(n|b)}^{\infty} e^{-t^{2}/2}\ dt,
\end{equation}

\noindent where $p(k|b) = \frac{b^{k}e^{-b}}{k!}$ is the Poisson
probability for observing $k$ events with $b$ events
expected. ${\mathcal S}(n|b)$ can be regarded as the number of standard
deviations required for $b$ to fluctuate to $n$, and it can be
calculated numerically. For $s+b$ expected events, the number of
observed events can be any number between $[0, \infty)$. The
significance is thus defined as:

\begin{equation}
\overline{{\mathcal S}} = \sum_{n=0}^{\infty} p(n|s+b) \cdot {\mathcal S}(n|b)
\label{eqn:significance}
\end{equation}

\noindent where $p(n|s+b)$ is the Poisson probability for observing $n$
events with $s+b$ events expected.

The NN output corresponding to the maximum significance determines our
cutoff to calculate the 95\% C.L. limit on the cross section. The error
on the expected signal includes uncertainties on trigger and object
identification efficiencies, on parton distribution functions (10\%),
differences between MCs (12\%), and on the jet energy scale
(5\%). Table~\ref{tbl:nn_result} lists the results in terms of 95\%
C.L. limits on production cross sections for various sets of model
parameters of mSUGRA.

\begin{table*}
\caption{Number of observed events ($N_{\rm obs}$), expected total
background events ($N^{\rm total}_{\rm bkgd}$), and expected signal
events ($N_{\rm mSUGRA}$), corresponding to the optimal NN cutoff for
different sets of mSUGRA parameters. The signal acceptance after NN
cutoff (Acceptance), mSUGRA production cross section for each parameter
set ($\sigma_{\rm mSUGRA}$), and the calculated 95\% C.L. upper limit on
the production cross section ($\sigma_{95\%}$) are also listed. All
limits are for $\tan\beta = 3$.}
\renewcommand{\baselinestretch}{1.2}
\begin{tabular}{cccccccc}
$m_{0}$ & $m_{1/2}$ & $N_{\rm obs}$ & $N^{\rm total}_{\rm bkgd}$ & $N_{\rm mSUGRA}$ & ${\rm Acceptance}$ & $\sigma_{\rm mSUGRA}$ & $\sigma_{95\%}$ \\
$(\rm{GeV})$ & $(\rm{GeV})$ & & & & (\%) & $(\rm{pb})$ & $(\rm{pb})$ \\
\tableline
$160$ & $60$ & $8$ & $6.45 \pm 1.22$ & $11.11 \pm 1.97$ & $0.360 \pm 0.064$ & $33.34$ & $29.61$ \\
$160$ & $65$ & $7$ & $5.94 \pm 1.15$ & $7.93 \pm 1.41$ & $0.364 \pm 0.065$ & $23.48$ & $26.87$ \\ \hline
$170$ & $58$ & $4$ & $4.43 \pm 0.88$ & $10.36 \pm 1.83$ & $0.301 \pm 0.053$ & $37.16$ & $23.59$ \\
$170$ & $65$ & $3$ & $2.87 \pm 0.61$ & $5.84 \pm 1.03$ & $0.283 \pm 0.050$ & $22.23$ & $23.71$ \\ \hline
$180$ & $60$ & $5$ & $4.18 \pm 0.85$ & $8.49 \pm 1.50$ & $0.305 \pm 0.054$ & $30.00$ & $27.76$ \\
$180$ & $67$ & $3$ & $3.45 \pm 0.72$ & $5.31 \pm 0.94$ & $0.306 \pm 0.054$ & $18.69$ & $20.89$ \\ \hline
$190$ & $55$ & $5$ & $5.51 \pm 1.12$ & $11.12 \pm 1.97$ & $0.248 \pm 0.044$ & $48.46$ & $30.88$ \\
$190$ & $63$ & $4$ & $3.65 \pm 0.79$ & $6.41 \pm 1.13$ & $0.299 \pm 0.053$ & $23.17$ & $25.15$ \\ \hline
$200$ & $57$ & $3$ & $2.72 \pm 0.60$ & $6.98 \pm 1.23$ & $0.208 \pm 0.037$ & $36.21$ & $32.79$ \\
$200$ & $62$ & $2$ & $2.31 \pm 0.51$ & $5.12 \pm 0.91$ & $0.231 \pm 0.041$ & $23.96$ & $24.85$ \\ \hline
$210$ & $53$ & $2$ & $2.75 \pm 0.59$ & $6.85 \pm 1.21$ & $0.096 \pm 0.017$ & $77.38$ & $57.99$ \\
$210$ & $60$ & $4$ & $3.74 \pm 0.81$ & $5.95 \pm 1.05$ & $0.238 \pm 0.042$ & $26.96$ & $31.33$ \\ \hline
$220$ & $50$ & $2$ & $3.72 \pm 0.79$ & $7.05 \pm 1.25$ & $0.054 \pm 0.009$ & $141.83$ & $97.55$ \\
$220$ & $55$ & $5$ & $4.02 \pm 0.83$ & $7.06 \pm 1.25$ & $0.169 \pm 0.030$ & $45.00$ & $50.87$ \\ \hline
$230$ & $45$ & $2$ & $2.90 \pm 0.62$ & $5.93 \pm 1.05$ & $0.030 \pm 0.005$ & $214.95$ & $183.99$ \\
$230$ & $50$ & $4$ & $3.45 \pm 0.74$ & $5.91 \pm 1.04$ & $0.046 \pm 0.008$ & $138.52$ & $166.06$ \\ \hline
$240$ & $43$ & $1$ & $2.53 \pm 0.56$ & $5.24 \pm 0.93$ & $0.023 \pm 0.004$ & $244.29$ & $194.22$ \\
$240$ & $52$ & $3$ & $3.83 \pm 0.80$ & $5.24 \pm 0.93$ & $0.056 \pm 0.010$ & $100.14$ & $110.68$ \\ \hline
$250$ & $41$ & $2$ & $3.47 \pm 0.72$ & $5.38 \pm 0.95$ & $0.021 \pm 0.004$ & $281.53$ & $256.82$ \\
$250$ & $42$ & $4$ & $4.97 \pm 0.96$ & $5.80 \pm 1.03$ & $0.024 \pm 0.004$ & $259.36$ & $282.43$ \\ \hline
$260$ & $41$ & $7$ & $5.91 \pm 1.16$ & $5.63 \pm 1.00$ & $0.022 \pm 0.004$ & $280.15$ & $452.28$ \\
$260$ & $42$ & $4$ & $3.87 \pm 0.77$ & $4.70 \pm 0.83$ & $0.020 \pm 0.003$ & $257.67$ & $374.37$ \\
\end{tabular}
\label{tbl:nn_result}
\end{table*}

\section{RESULTS} \label{sec:results}

\begin{figure}
        \epsfig{file=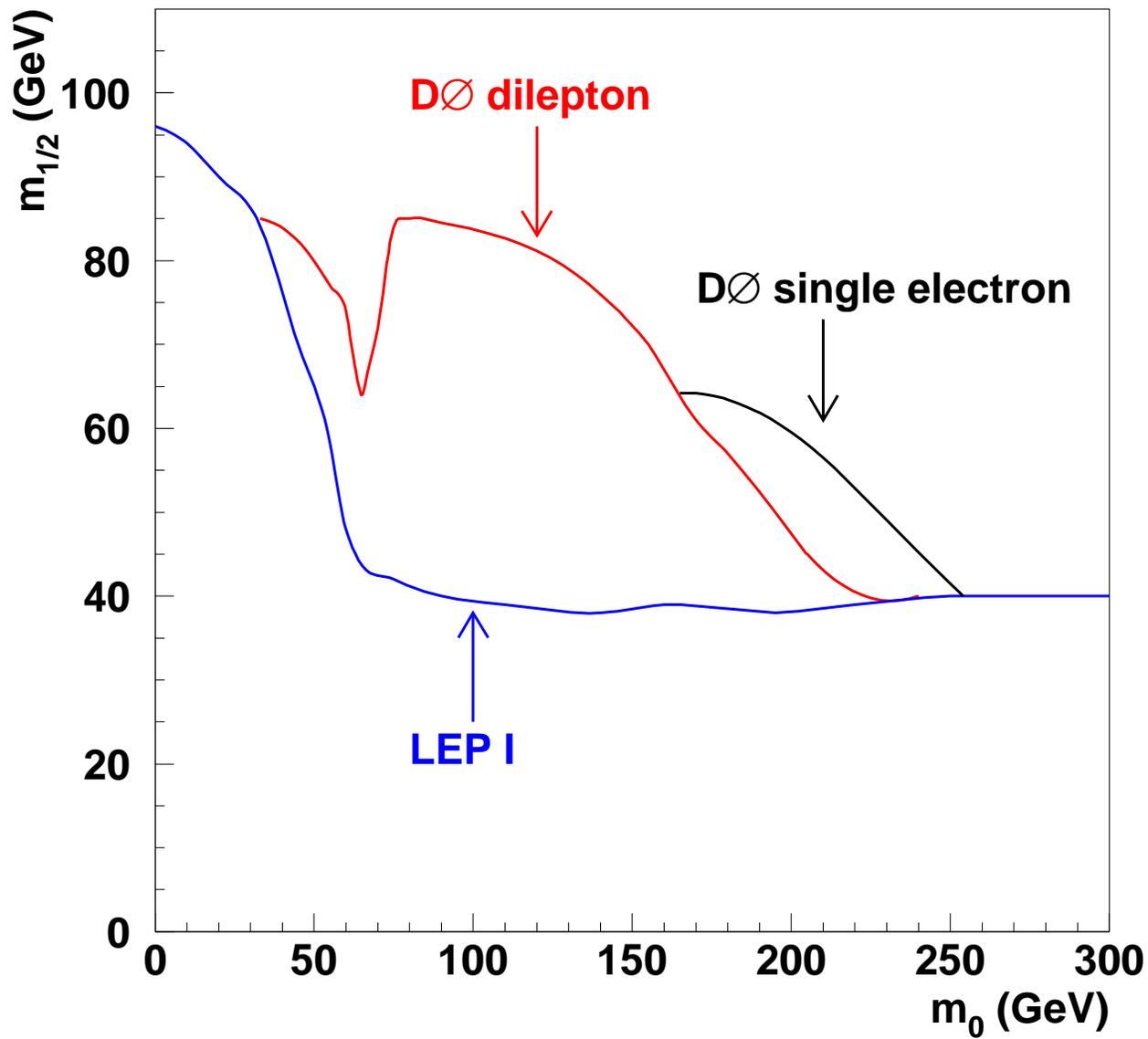, width=\hsize}
        \caption{Exclusion contour at the 95\% C.L. for mSUGRA with
        $\tan\beta = 3$. The result from the D\O\ dilepton and LEP I
        analyses are also shown.}
	\label{fig:tb3_excl}
\end{figure}

We conduct an independent NN analysis on each generated mSUGRA
point. The production cross section calculated by {\sc pythia} is
compared with that obtained by limit calculation at 95\% C.L. to
determine whether the mSUGRA point is excluded or not. Using the two
cross sections at each point, we linearly extrapolate between the
excluded and non-excluded points to determine the exact location of the
exclusion contour. The exclusion contour at the 95\% C.L. is plotted in
Fig.~\ref{fig:tb3_excl}. Shown in the same figure are the results of the
D\O\ dilepton and LEP I~\cite{lepsusy} analyses.

Our single-electron analysis is particularly sensitive in the moderate
$m_{0}$ region. The extended region of exclusion relative to the D\O\
dilepton result is in the range of $165 \; {\rm GeV} < m_{0} < 250 \;
{\rm GeV}$. The dominant SUSY process changes from \gluino\-\sq\
production at $m_{0}=170 \; {\rm GeV}$ to \gluino\ pair production at
$m_{0}=250 \; {\rm GeV}$. The limit worsens as $m_{0}$ increases because
the mass difference between \charginoa\ (\neutralinob) and
\neutralinoa\ decreases, resulting in softer electron and jets spectra,
and consequently reduced acceptance.

As this work was being completed, a related
result~\cite{cdf_jets_met_prl} on searches for mSUGRA in the jets plus
missing energy channel at Tevatron appeared. Since its limits on mSUGRA
parameters, although more restrictive than those obtained in this work
and in the earlier D\O\ publication~\cite{d0_jets_met_prl} in the
analogous channel, are expressed in a different parameter plane
($m_{\widetilde{q}}$ vs. $m_{\widetilde{g}}$), we do not show them in
Fig.~\ref{fig:tb3_excl}.

\section{CONCLUSION} \label{sec:conclusion}

We observe 72 candidate events for an mSUGRA signal in the final state
containing one electron, four or more jets, and large \MET\ in $92.7 \;
\rm{pb}^{-1}$ data. We expect $80.3 \pm 10.4$ such events from
misidentified multijet, \ttbar, $W+$~jets, and $WW$ production. We
conclude that there is no evidence for the existence of mSUGRA. We use
neural network to select a kinematic region where signal to background
significance is the largest. The upper limit on the cross section
extends the previously D\O\ obtained exclusion region of mSUGRA
parameter space.

\section*{Acknowledgments}
%
We thank the staffs at Fermilab and collaborating institutions, 
and acknowledge support from the 
Department of Energy and National Science Foundation (USA),  
Commissariat  \` a L'Energie Atomique and 
CNRS/Institut National de Physique Nucl\'eaire et 
de Physique des Particules (France), 
Ministry for Science and Technology and Ministry for Atomic 
   Energy (Russia),
CAPES and CNPq (Brazil),
Departments of Atomic Energy and Science and Education (India),
Colciencias (Colombia),
CONACyT (Mexico),
Ministry of Education and KOSEF (Korea),
CONICET and UBACyT (Argentina),
The Foundation for Fundamental Research on Matter (The Netherlands),
PPARC (United Kingdom),
Ministry of Education (Czech Republic),
A.P.~Sloan Foundation,
NATO, and the Research Corporation.


\begin{thebibliography}{99}
\bibitem{eta_definition} At D\O, we define $\theta$ and $\phi$
to be the polar and azimuthal angles of the physical objects,
respectively. We define the pseudorapidity $\eta = -\ln(\tan\theta/2)$.
We denote $\eta_{\rm d}$ as the pseudorapidity calculated relative to
the center of the detector rather than relative to the reconstructed
interaction vertex.
\end{thebibliography}
\end{document}